\documentclass[11pt]{article}
\usepackage{jheppub}
\usepackage{amsmath, amsfonts, amssymb, array}
\usepackage{mathrsfs}
\usepackage{slashed}
\newcolumntype{C}[1]{>{\centering\let\newline\\\arraybackslash\hspace{0pt}}m{#1}}
\makeatletter
\g@addto@macro\bfseries{\boldmath}
\makeatother
\usepackage[english]{babel}
\usepackage[autostyle]{csquotes}
\usepackage{float}

\newcommand{\be}{\begin{equation}}
\newcommand{\ee}{\end{equation}}
\newcommand{\bea}{\begin{eqnarray}}
\newcommand{\eea}{\end{eqnarray}}
\newcommand{\nn}{\nonumber}

\usepackage{hyperref}

\def\a{\alpha}
\def\b{\beta}
\def\g{\gamma}

\def\d{\delta}

\def\m{\mu}
\def\n{\nu}
\def\r{\rho}
\def\s{\sigma}

\def\o{\omega}

\def\nn{\nonumber}

\def\nn{\nonumber}

\newcommand{\ben}[1]{\begin{eqnarray}\label{#1} }
\newcommand{\een}{\end{eqnarray}}

\newcommand{\DD}{{\cal D}}

\newcommand{\cA}{{\cal A}}
\newcommand{\thc}{\text{h.c.}}
\newcommand{\Dslash}{\not{\hbox{\kern-4pt $D$}}}
\newcommand{\pslash}{\not{\hbox{\kern-4pt $\partial$}}}
\newcommand{\Dcslash}{\not{\hbox{\kern-4pt $\DD$}}}

\usepackage{graphicx}
\usepackage{amssymb}
\usepackage{amsthm}
\usepackage{amsmath}
\usepackage{slashed}
\usepackage{color}
\usepackage{bbold}

\numberwithin{equation}{section}

\newcommand {\cB}{{\cal B}}
\newcommand {\cC}{{\cal C}}
\newcommand {\cD}{{\cal D}}

\newcommand {\cF}{{\cal F}}
\newcommand {\cG}{{\cal G}}
\newcommand {\cH}{{\cal H}}

\newcommand {\cL}{{\cal L}}

\newcommand {\cP}{{\cal P}}

\newcommand {\cR}{{\cal R}}

\newcommand {\cV}{{\cal V}}

\newcommand {\cZ}{{\cal Z}}

\def\a{\alpha}
\def\b{\beta}

\def\d{\delta}

\def\g{\gamma}

\def\m{\mu}
\def\n{\nu}
\def\o{\omega}

\def\r{\rho}
\def\s{\sigma}

\makeatletter
\g@addto@macro\bfseries{\boldmath}
\makeatother

\title{Curvature squared action in four dimensional $N=2$ supergravity using the dilaton Weyl multiplet}

\author{Madhu Mishra}
\author{and Bindusar Sahoo}

\affiliation{
	Indian Institute of Science Education and Research,
	Vithura, Thiruvananthapuram - 695551, India}

\emailAdd{madhu50315@iisertvm.ac.in}
\emailAdd{bsahoo@iisertvm.ac.in}

\abstract{In this paper we derive the most general curvature squared action coupled to an arbitrary number of vector multiplets in four dimensional $N=2$ supergravity using the dilaton Weyl multiplet. The action that we derive is encoded in a single holomorphic prepotential.
}

\begin{document}
	\allowdisplaybreaks
	\maketitle

	\section{Introduction}
	
	The fascination for studying higher derivative invariants arose in the mid-1970s, with the hope of the cancellation of ultraviolet (UV) divergence in supergravity. Although  pure supergravity theories with ${N}= 1, 2$
	are UV finite at both one and two loops \cite{PhysRevLett.38.527} ; the finiteness fails for three loops.
	Nonetheless, UV cancellation for higher loops has been expected in  case of ${N} > 2$. From the perspective of superstring theory, it has been suggested that the maximal supergravity i.e. ${N}= 8$ supergravity in four dimensions might be UV finite up to eight loops \cite{PhysRevLett.98.131602}. This hope is boosted up by the fact that maximally supersymmetric Yang-Mills theory i.e. ${N}= 4$ super-Yang Mills in four dimensions is UV finite \cite{Brink:1982wv}. Other than serving as a candidate counter-term for cancelling the UV divergences  in supergravity theories, the construction of higher derivative invariants is also crucial to obtain the finite sub-leading corrections to the black hole entropy.
	In the large charge limit the dynamics of black holes is described by two-derivative action and entropy of such black holes is given by the Bekenstein-Hawking area law. However, for a finite sized black hole, curvature and field strengths are prominent and can not be neglected at higher orders. Hence, higher derivative terms are required to describe the geometry of a finite sized black hole which leads to the sub-leading corrections to the black hole entropy. 
	
	\noindent
	It is well known that $R^2$ - supergravity naturally arises in ten dimensions when one considers the low energy limit of superstring theories. More importantly, we have the quantum description of black holes in both theories. In supergravity theories, black holes are the supersymmetric analogues of the classical black holes, and their quantum behavior is much constrained due to additional supersymmetry. On the other hand, black holes in string theory are black objects generally realized as multiple configuration of D-branes, NS5-branes and fundamental strings wrapping some cycles of the compact internal manifold together with fluxes. There are several such configurations that can give rise to the same black hole with the same asymptotic charges and each such configuration is referred to as the microstate of the black hole. One of the consistency checks of string theory as a mathematically self consistent theory of quantum gravity is the matching of the degeneracy of such microstates with the macroscopic entropy computed from the underlying low energy supergravity. In the large charge limit, one finds the agreement between the macroscopic Bekenstein-Hawking entropy and microscopic entropy in string theory for extremal charged/rotating black holes \cite{ Strominger:1996sh, Kaplan:1996ev}. For deviation from the large charge limit, one needs to go beyond the standard two derivative theory of gravity and consider higher derivative corrections to the Bekenstein-Hawking entropy. Such higher derivative corrections has been extensively studied  and are shown to match the microscopic degeneracy for a large class of extremal BPS black holes \cite{LopesCardoso:1998tkj,Sen:2007qy,Sahoo:2006pm,Sahoo:2006rp,Prester:2008iu,Cvitan:2007hu,Banerjee:2016qvj,Mohaupt:2000mj}.
	
	\noindent
	As it is clear from the above mentioned discussions, higher derivative corrections in supergravity are important. For several purpose one needs to construct the higher derivative corrections to supergravity in an off-shell manner. One of the reason is technical. Since in the off-shell construction one does not need to change the supersymmetry transformations of the component fields, the problem becomes more tractable. The other motivation comes from the use of quantum entropy function \cite{Sen:2008vm} and localization techniques in the computation of exact quantum entropy of black holes \cite{Dabholkar:2011ec,Dabholkar:2010uh} which relies on some of the supersymmetries to be realized off-shell. The off-shell construction of higher derivative supergravity is facilitated by the use of the superconformal methods. In this method, one first constructs a theory of superconformal gravity. In an N-extended superconformal gravity which has N ordinary supersymmetries, one also has additional symmetries like the dilatation, special conformal transformation, R-symmetries and S-supersymmetry or special-supersymmetry. The degrees of freedom in an N-extended superconformal gravity are arranged in several multiplets. The Weyl multiplet contains the graviton, its supersymmetric partner gravitino and other auxiliary fields needed for the supersymmetric completion of the multiplet. There are several matter multiplets: for example a vector multiplet which contains a vector gauge field or a tensor multiplet which contains a tensor gauge field. Some of these matter multiplets are used as compensators for gauge fixing the additional symmetries present in superconformal gravity and ultimately becomes a part of the gravity multiplet in the gauge fixed theory. This gauge fixed theory is nothing but the ordinary supergravity with super-Poincar{\'e} symmetries which we will refer to as super-Poincar{\'e} gravity. Depending on the details of the construction, the resulting super-Poincar{\'e} gravity will typically have higher derivative corrections. Using this method a large class of higher derivative invariants have been constructed in N=2 supergravity in four dimensions \cite{Bergshoeff:1980is,deWit:2010za,Butter:2013lta,Kuzenko:2015jxa}.
	
	\noindent
	The Weyl multiplet that contains the graviton typically comes in two variants: the standard Weyl multiplet and the dilaton Weyl multiplet. They differ from each other in their auxiliary fields. The dilaton Weyl multiplet has a scalar field with mass dimension $+1$ and the multiplet derives its name from this scalar field which can be interpreted as a dilaton. The dilaton Weyl multiplet was known for the minimal conformal supergravity in five and six dimensions since long \cite{BERGSHOEFF1986653, Bergshoeff:2001hc} and was recently discovered in $N=2$ conformal supergravity four dimensions \cite{Butter:2017pbp}. The dilaton Weyl multiplet has further been used to construct all curvature squared invariants in five dimensional and six dimensional minimal supergravity \cite{Ozkan:2013nwa,Butter:2018wss}. In four dimensional $N=2$ supergravity, the Weyl squared invariant has long been known based on the construction of $N=2$ conformal supergravity \cite{Bergshoeff:1980is} using standard Weyl multiplet. A second class of invariant which is a particular combination of Ricci scalar square and Ricci tensor square was obtained in \cite{Butter:2013lta} using the logarithm of a conformal primary chiral superfield. A particular combination of this invariant together with the Weyl squared invariant gives rise to the Gauss Bonnet invariant. The third curvature squared invariant was constructed using the tensor multiplet \cite{Kuzenko:2015jxa} and gives rise to the Ricci scalar squared invariant. All the above mentioned constructions in four dimensional $N=2$ supergravity made use of the standard Weyl multiplet and together they complete the all curvature squared invariant in $N=2$ supergravity in four dimensions using the standard Weyl multiplet
	
	\noindent
	In this paper, we would like to construct all curvature squared invariants coupled to an arbitrary number of vector multiplets in four dimensional $N=2$ supergravity using the dilaton Weyl multiplet . One of the advantages of using a dilaton Weyl multiplet is that only one compensating multiplet is needed to go from the superconformal theory to the super-Poincar{\'e} theory whereas standard Weyl multiplet needs two compensating multiplets. The reason is as follows. In the standard Weyl multiplet the auxiliary scalar field $D$ which is a fundamental field belonging to the Weyl multiplet appears as a Lagrange multiplier and hence it must be cancelled by another scalar degree of freedom coming from a different compensating multiplet. Similar situation arises with the field $\chi_i$ in the fermionic sector. This is the reason why an additional compensator is required while working with the standard Weyl multiplet. Whereas in dilaton Weyl multiplet there is no fundamental auxiliary field $D$ or $\chi_i$ in the multiplet and hence such problems do not arise. Hence one can construct pure $N=2$ supergravity with fewer off-shell degrees of freedom using the dilaton Weyl multiplet vis-{\`a}-vis a standard Weyl multiplet. Another advantage of using the dilaton Weyl multiplet is that it naturally comes with a two form field whereas there is no two form field in the standard Weyl multiplet. Any higher derivative action that we would like to construct that might come from the compactification of string theory needs to have a two form field. While the dilaton Weyl multiplet has a two form field inbuilt, the two form field in a construction involving standard Weyl multiplet must come from another matter multiplet. 
	
	\noindent
	The construction of Weyl squared and the Ricci scalar squared invariant closely follows the construction using the standard Weyl multiplet. The construction of Riemann squared invariant is made possible due to the map between a Yang-Mills multiplet and the dilaton Weyl multiplet once we gauge fix the superconformal theory to a super-Poincar{\'e} one. Such maps were known to exist in six dimensions \cite{BERGSHOEFF198673} and five dimensions \cite{Ozkan:2013uk,Ozkan:2013nwa}. In this paper we construct this map in four dimensions which allows us to construct the Riemann squared invariant in four dimensional $N=2$ supergravity. We club together the construction of all three curvature squared invariants and its coupling to an arbitrary number of vector multiplets encoded in a single prepotential $G(X^I,B,\cA,\cP)$ which is an arbitrary holomorphic function of the vector multiplet scalars and the chiral background constructed out of the tensor multiplet, the Yang-Mills multiplet and the dilaton Weyl multiplet. 
	
\noindent
	This paper is organized as follows. In section-\ref{2}, we briefly review $N=2$ conformal supergravity and the several multiplets that we would be using. In section-\ref{3}, we will review the construction of superconformal actions using standard density formulae such as the tensor-vector density formula and the chiral density formula. We will review the construction of the superconformal action for the Yang-Mills multiplet \cite{deWit:1980lyi} and the tensor multiplet \cite{deWit:1982na} using the above mentioned density formulae. In section-\ref{4}, we will discuss super-Poincar{\'e} gauge fixing and find the transformation rules for the dilaton Weyl multiplet in the gauge fixed theory which would be used later. In section-\ref{sec5}, we will use the gauge fixing condition derived earlier to find the action of the tensor multiplet in the super-Poincar{\'e} background. This will give rise to the Einstein-Hilbert term in the Jordan frame. We also establish the map between the Yang-Mills multiplet and the dilaton Weyl multiplet. Furthermore, we use the map to construct the off-shell supersymmetric Riemann tensor squared action. In section-\ref{6}, we generalize the construction of the Riemann squared invariant in section-\ref{sec5} to the most general $N=2$ Poincar{\'e} supergravity with arbitrary curvature squared corrections and coupled to an arbitrary number of vector multiplets encoded in a single holomorphic prepotential.  In section-\ref{conc}, we summarize our results and end with some future directions. We have three appendices. In appendix-\ref{A1} we discuss the notations that we have used for the covariant derivatives in our paper. In appendix-\ref{A}, we derive several relations concerning the prepotential and the derivatives with respect to its arguments. In appendix-\ref{SU2breaking}, we discuss the gauge fixing of $SU(2)$-R symmetry. 
	
	\section{$N=2$ superconformal gravity in D=4}\label{2}
	A theory of conformal supergravity with an $N$-extended supersymmetry in four dimensions is a gauge theory based on the superconformal algebra $su(2,2|N)$ with appropriate set of curvature constraints that makes the gauge theory into a theory of gravity. The multiplet of fields that contains the gauge fields is also known as the Weyl multiplet. For $N>1$, one also needs to add extra set of auxiliary fields to the multiplet as well as modify the superconformal algebra to a soft superconformal algebra where the structure constants becomes structure functions depending on the auxiliary fields. There are a couple of different ways to select the auxiliary fields which leads to different Weyl multiplets: the standard and the dilaton Weyl multiplet which has the same gauge field contents but different auxiliary field contents. The different types of Weyl multiplets were already known for five and six spacetime dimensions \cite{BERGSHOEFF1986653, Bergshoeff:2001hc} whereas for four dimensions, the dilaton Weyl multiplet was constructed recently in \cite{Butter:2017pbp}. The dilaton Weyl multiplet plays a crucial role in our construction and we will discuss its structure in the following subsection.
	\subsection{$N=2$ Dilaton Weyl Multiplet }
	The independent field contents of the dilaton Weyl multiplet are tabulated in Table-\ref{Table-dilatonWeyl}, where we have also given their Weyl weight, chiral weight as well as chirality for fermions:
	\begin{table}[H]
		\caption{N=2 superconformal dilaton Weyl multiplet}\label{Table-dilatonWeyl}
		\begin{center}
			\begin{tabular}{ | p{1cm}|p{2cm}|p{1cm}|p{1cm}|p{2cm}| }
				\hline
				Field & SU(2) Irreps & Weyl weight (w) & Chiral weight (c) & Chirality fermions \\ \hline
				$e_{\mu}{}^{a}$ & $\bf{1}$ & -1 & 0 & -- \\ \hline
				$V_{\mu}{}^{i}{}_{j}$ & $\bf{3}$ & 0 & 0 & -- \\ \hline
				$b_{\mu}$ & $\bf{1}$ & 0 & 0 & -- \\ \hline
				$X$ & $\bf{1}$ & 1 & -1 & -- \\ \hline
				$W_{\mu}$ & $\bf{1}$ & 0 & 0 & -- \\ \hline
				$\tilde{W}_{\mu}$ & $\bf{1}$ & 0 & 0&-- \\ \hline
				$B_{\mu\nu}$ & $\bf{1}$ & 0 & 0 & -- \\ \hline
				$\psi_{\mu}{}^{i}$ & $\bf{2}$ & -1/2 & -1/2 & +1 \\ \hline
				$\Omega_{i}$ & $\bf{2}$ & 3/2 & -1/2 & +1 \\ \hline
			\end{tabular}
		\end{center}
	\end{table}
	\noindent
	There are additional U(1) gauge symmetries corresponding to the U(1) gauge fields $W_{\mu}$ and $\tilde{W}_{\mu}$ and there is an additional tensor gauge symmetry corresponding to the tensor gauge field $B_{\mu\nu}$. Transformation of the independent fields under all the superconformal transformations and the above mentioned gauge transformations are given below. The parameters corresponding to the different gauge transformations is tabulated in Table-\ref{Table-parameters}. We will be referring to the ordinary supersymmetry as Q-supersymmetry and the special supersymmetry present in conformal supergravity as S-supersymmetry.
	\begin{align}\label{dwf}
	\delta e_{\mu}{}^{a}&=\bar{\epsilon}^{i}\gamma^{a}\psi_{\mu i}+\thc-\Lambda_D e_{\mu}{}^{a}+\Lambda_{M}^{ab}e_{\mu b} ,\nonumber \\
	\delta\psi_{\mu}{}^{i}&=2\mathcal{D}_{\mu}\epsilon^{i}-\frac{1}{8}\varepsilon^{ij}\bar{X}^{-1} \gamma\cdot (\mathcal{F}^{-} + i \mathcal{G}^-)\gamma_{\mu}\epsilon_{j}-\gamma_{\mu}\eta^{i}-\frac{1}{2}\Lambda_D\psi_{\mu}{}^{i}\nn\\
	&\quad -\frac{i}{2}\Lambda_A\psi_{\mu}{}^{i}+\Lambda^{i}{}_{j}\psi_{\mu}{}^{j} +\frac{1}{4}\Lambda_{M}^{ab}\gamma_{ab}\psi_{\mu}{}^{i}\nonumber \\
	\delta b_{\mu}&=\frac{1}{2}\bar{\epsilon}^{i}\phi_{\mu i}-\frac{1}{4}X^{-1}\bar{\epsilon}^{i}\gamma_{\mu}\Dslash\Omega_{i}-\frac{1}{2}\bar{\eta}^{i}\psi_{\mu i}+\thc+\Lambda_{K}^{a}e_{\mu a} +\partial_{\mu}\Lambda_D\nonumber \\
	\delta \mathcal{V}_{\mu}{}^{i}{}_{j}&=2\bar{\epsilon}_{j}\phi_{\mu}^{i}-\bar{X}^{-1}\bar{\epsilon}_{j}\gamma_{\mu}\Dslash\Omega^{i}+2\bar{\eta}_{j}\psi_{\mu}^{i}-(\thc; ~\text{traceless})-2\partial_{\mu}\Lambda^{i}{}_{j} \nonumber \\ &\quad +\Lambda^{i}{}_{k}\mathcal{V}_{\mu}{}^{k}{}_{j}-\Lambda^{k}{}_{j}\mathcal{V}_{\mu}{}^{i}{}_{k}\nonumber \\
	\delta X &= \bar{\epsilon}^{i}\Omega_{i}+\left(\Lambda_D -i\Lambda_A\right)X \nonumber \\
	\delta\Omega_{i}&=2\Dslash X \epsilon_{i}+\frac{1}{4}\varepsilon_{ij}\gamma\cdot {\mathcal{F}}\epsilon^{j}-\frac{i}{4}\varepsilon_{ij}\gamma\cdot \mathcal{G}^{-}\epsilon^{j}+2X\eta_{i} +\left(\frac{3}{2}\Lambda_D -\frac{i}{2}\Lambda_A\right)\Omega_{i}\nonumber \\
	&\quad -\Lambda^{j}{}_{i}\Omega_{j}+\frac{1}{4}\Lambda_{M}^{ab}\gamma_{ab}\Omega_{i}\nonumber \\
	\delta W_{\mu}&=\varepsilon^{ij}\bar{\epsilon}_{i}\gamma_{\mu}\Omega_{j}+2\varepsilon_{ij}\bar{X}\bar{\epsilon}^{i}\psi_{\mu}^{j}+\thc+\partial_{\mu}\lambda\nonumber \\
	\delta \tilde{W}_{\mu}&=i\varepsilon^{ij}\bar{\epsilon}_{i}\gamma_{\mu}\Omega_{j}-2i\varepsilon_{ij}\bar{X}\bar{\epsilon}^{i}\psi_{\mu}^{j}+\thc+\partial_{\mu}\tilde{\lambda}
	\nonumber\\
	\delta B_{\mu\nu}&=\tfrac12 W_{[\mu} \delta_Q W_{\nu]}+\tfrac12 \tilde{W}_{[\mu} \delta_Q \tilde{W}_{\nu]}+\bar X \bar\epsilon^i\gamma_{\mu\nu}\Omega_i+ X\bar\epsilon_i\gamma_{\mu\nu} \Omega^i+2\,X\bar X\bar\epsilon^i\gamma_{[\mu}\psi_{\nu]\,i} \nonumber \\
	&\quad +2\,X\bar X \bar\epsilon_i\gamma_{[\mu}\psi_{\nu]}^i+2\partial_{[\mu}\Lambda_{\nu]}-\frac{\lambda}{4}{F}_{\mu\nu}-\frac{\tilde{\lambda}}{4}{G}_{\mu\nu}
	\end{align}
	where
	\begin{align}\label{delmuepsilon}
	\mathcal{D}_{\mu}\epsilon^{i} = 
	\partial_{\mu}\epsilon^{i}
	-\frac{1}{4}\omega_{\mu}{}^{ab}\gamma_{ab}\epsilon^{i}
	+\frac{1}{2}\left(b_{\mu}+iA_{\mu}\right)\epsilon^{i}
	+\frac{1}{2}\mathcal{V}_{\mu}{}^{i}{}_{j}\epsilon^{j}~.
	\end{align}
\begin{table}[H]
		\caption{Superconformal symmetries and their transformation parameters.}\label{Table-parameters}
		\begin{center}
			\begin{tabular}{ | p{1cm}|p{1cm}|p{1cm}|p{1.5cm}|p{1.5cm}| p{1cm}|p{1cm}|p{1cm}|p{1cm}|p{1cm}|}
				\hline
				Transf: & Q-susy & S-susy & Lorentz & Dilatation & $U(1)$ R-symm & $SU(2)$ R-symm & $U(1)_W$ & $U(1)_{\tilde{W}}$& Tensor gauge\\ \hline
				Param-eters: &\; $\epsilon$ & \;$\eta$ & \;$\Lambda_{M}^{ab}$ &\; $\Lambda_{D}$ & \;$\Lambda_{A}$ & \;$\Lambda^{i}{}_{j}$ &\; $\lambda$ & \;$\tilde{\lambda}$ & \;$\Lambda_{\mu}$ \\ \hline
			\end{tabular}
		\end{center}
	\end{table}	
	
	
	
	\noindent
	In the above equation, $\mathcal{F}_{\mu\nu}$ and $\mathcal{G}_{\m \n}$ are the superconformally covariant field strength associated with the gauge fields $W_\mu$ and $\tilde{W}_{\mu}$ respectively.
	\begin{align}\label{Bfieldstrength}
	\mathcal{F}_{\mu\nu}&=2\partial_{[\mu}W_{\nu]}-\varepsilon_{ij}\bar{\psi}_{[\mu}{}^{i}\Big(\gamma_{\nu]}\Omega^{j}+\psi_{\nu]}{}^j \bar X\Big)-\varepsilon^{ij}\bar{\psi}_{[\mu i}\Big(\gamma_{\nu]}\Omega_{j}+\psi_{\nu]j} X\Big)\;.\nonumber\\
	\mathcal{G}_{\mu\nu}&=2\partial_{[\mu}\tilde{W}_{\nu]}+i\varepsilon_{ij}\bar{\psi}_{[\mu}{}^{i}\Big(\gamma_{\nu]}\Omega^{j}+\psi_{\nu]}{}^j \bar X\Big)-i\varepsilon^{ij}\bar{\psi}_{[\mu i}\Big(\gamma_{\nu]}\Omega_{j}+\psi_{\nu]j} X\Big)
	\end{align}
	\noindent
	The superconformally covariant field strength $\cH_{\mu\nu\rho}$ associated with the two-form gauge fields $B_{\mu \n}$ is given as
	\begin{eqnarray}\label{3formfieldstrength}
	\cH_{\mu\nu\rho} &=& 3\partial_{[\mu} B_{\nu\rho]}+\tfrac34 W_{[\mu}F_{\nu\rho]}+\tfrac34 \tilde{W}_{[\mu}G_{\nu\rho]} \nonumber \\ &&- \tfrac32\bar X\bar\psi_{[\mu}^i\gamma_{\nu\rho]}\Omega_i-\tfrac32 X \bar\psi_{[\mu\,i}\gamma_{\nu\rho]}\Omega^i-3 X\bar X \bar\psi_{[\mu}^i\gamma_\nu\psi_{\rho]}{}_i~,
	\end{eqnarray}
	
	
	\noindent
	Apart from the independent fields tabulated in Table-\ref{Table-dilatonWeyl}, the dilaton Weyl multiplet has some dependent gauge fields: $\omega_{\mu}{}^{ab}$, $f_{\mu}^{a}$, $\phi_{\mu}^{i}$ and $A_{\mu}$ corresponding to local Lorentz transformation, special conformal transformations, S-supersymmetry and U(1)$_R$ symmetry respectively. The dependent fields are obtained by solving the following constraints:
	\begin{eqnarray}\label{constraints}
	R(P)_{\mu\nu}&=&0\\
	\gamma^{\mu}{R}(Q)_{\mu\nu}&=&0\\
	e_{b}{}^{\nu}{R}(M)_{\mu\nu}{}_{a}{}^{b}-i\tilde{{R}}(A)_{\mu a}&=&0\\
	X\bar{X}D_{a}\log\left(\frac{X}{\bar{X}}\right)&=&\frac{1}{2}\bar{\Omega}_{k}\gamma_{a}\Omega^{k}+\frac{1}{6}\varepsilon_{abcd}\mathcal{H}^{bcd}\;,
	\end{eqnarray}
	where $R(P)_{\mu\nu}$, ${R}(Q)_{\mu\nu}$, ${R}(M)_{\mu\nu}{}^{ab}$, $R(A)_{\mu\nu}$ are respectively the fully supercovariant curvatures for local translation, Q-supersymmetry, local Lorentz transformations and U(1) R-symmetry. $\tilde{R}(A)$ denotes the dual of $R(A)$. $\mathcal{H}^{bcd}$ is the fully supercovariant field strength of the tensor gauge field $B_{\mu\nu}$ and $D_{a}$ is used to denote the fully supercovariant derivative. The above constraints can be solved and one can obtain the expressions for the dependent gauge fields in terms of the independent ones as shown below.
	\begin{eqnarray}\label{dependent}
	\omega_{\mu}^{ab}&=&-2e^{\nu[a}\partial_{[\mu}e_{\nu]}{}^{b]}-e^{\nu[a}e^{b]\sigma}e_{\mu c}\partial_{\sigma}e_{\nu}{}^{c}-2e_{\mu}{}^{[a}e^{b]\nu}b_{\nu}-\frac{1}{4}\left(2\bar{\psi}_{\mu}^{i}\gamma^{[a}\psi_{i}^{b]}+\bar{\psi}^{ai}\gamma_{\mu}\psi_{i}^{b}+\text{h.c}\right)\nonumber \\
	A_{a}&=& -\tfrac i{12}\tfrac{1}{X\bar{X}}\varepsilon_{abcd}\,\cH^{bcd}+\tfrac{i}{2}\partial_a \log \frac{X}{\bar X}-\tfrac{i}{4X}\bar{\psi}_{a}^{i}\Omega_{i}+\tfrac{i}{4\bar{X}}\bar{\psi}_{ai}\Omega^{i}-\tfrac i{4}\tfrac{1}{X\bar{X}}\bar{\Omega}^{k}\gamma_{a}\Omega_{k}\nonumber \\
	\phi_{\m}{}^{i} &=& \frac{1}{4} \left( \gamma^{\r \s} \gamma_\m - \frac{1}{3} \gamma_\m \gamma^{\r \s} \right)( {R}'(Q)_{\r \s}^{i} + iA_{[\r}\psi_{\s]}{}^i) \nonumber\\
	f_{\m}{}^{a} &=& \frac{1}{2} {R}_{\m}{}^{a} - \frac{1}{12}{R}\; e_{\m}{}^a -\frac{1}{2}i \tilde{R}_\m{}^a (A)\;,
	\end{eqnarray}
	where $ {R}'^{i} _{\m \n}(Q)$ is obtained from the fully supercovariant curvature of Q-supersymmetry $R(Q)^{i}_{\mu\nu}$ by subtracting the S-supersymmetry and U(1) R-symmetry  covariantization terms and $R_{\mu}{}^{a}$ is obtained from the contraction of the fully supercovariant curvature $R(M)_{\mu\nu}{}^{ab}$ of local Lorentz transformations after subtracting the $f_{\mu}^{a}$ term. $R$ is defined as the contraction of $R_{\mu}{}^{a}$. At bosonic level $R_{\mu}{}^{a}$ and $R$ coincides with the standard definition of Ricci tensor and Ricci scalar respectively once Poincar{\'e} gauge fixing condition $b_{\mu}=0$ is imposed (See section-\ref{4} for the Poincar{\'e} gauge fixing). See the following equations for the definitions. 
	\bea\label{supcurv}
	{R}_{\m \n}(Q)^i &=& {R}_{\m \n}'(Q)^{i} +i A_{[\m}\psi_{\n]}{}^i - \gamma_{[\m}\phi_{\n]}{}^i   ,\nonumber\\ 
	{R}_{\m \n}'(Q)^{i} &=& 2 \mathcal{D}'_{[\m}\psi_{\n]}{}^i -  \frac{1}{16} \varepsilon^{ij}\gamma^{a b} T_{a b}^{-} \gamma_{[\m} \psi_{\n] j}\;,\nn\\
	R_{\mu}^{a}&=&\left(R(M)_{\mu\nu}{}^{ab}e^{\nu}_{b}\right)_{f=0}\;,\; R=R_{\mu}^{a}e^{\mu}_{a}\;,\nn\\
	\text{where}\; R(M)_{\mu\nu}{}^{ab}&=&2\partial_{[\mu}\omega_{\nu]}^{ab}-2\omega_{[\mu}^{ac}\omega_{\nu]c}{}^{b}-4f_{[\mu}{}^{[a}e_{\nu]}{}^{b]}+\text{fermions}\;,
	\eea
	where, $\mathcal{D}'_{\mu}$ is used to denote the fully supercovariant derivative w.r.t all bosonic gauge transformations except U(1) R-symmetry and special conformal transformations. Wherever required, we will use $\mathcal{D}_{\mu}$ without the prime to denote the fully supercovariant derivative w.r.t all the bosonic gauge transformations except special conformal transformations (See Appendix-\ref{A1} for our notations with covariant derivatives). One can take some composite expressions of the dilaton Weyl multiplet fields as shown below which transforms like the auxiliary fields of a standard Weyl multiplet.
	\begin{eqnarray}\label{T}
	{T}^{+ab} &=&2\,X^{-1}\left(\mathcal{F}^{+ab}-i\mathcal{G}^{+ab}\right), \quad {T}^{-ab}=2\,\bar{X}^{-1}\left(\mathcal{F}^{-ab}+i\mathcal{G}^{-ab}\right),\nonumber\\
	{\chi}_{j}&=&\frac{1}{2}X^{-1}\Dslash\Omega_{j},
	\nonumber\\
	{D} &=& \frac{1}{2}|X|^{-2}\left(X\square^c \bar{X}+\frac{1}{2}\bar{\Omega}^{k}\Dslash \Omega_{k}+\frac{1}{4}\mathcal{F}\cdot\mathcal{F}^{+}+\frac{1}{4}\mathcal{G}\cdot\mathcal{G}^{+}+\thc\right)
	\end{eqnarray}
	where $\square^c{X}$ is the fully superconformal de'Alembertian as shown below. 
	\begin{eqnarray}
	\square ^{c}{X} &=&  \mathcal{D}^a D_a X - \frac{1}{2} \bar{\psi}_{a}{}^{i} D^a \Omega_{i} +  f_a^a X   - \frac{1}{4} \bar{\Omega}_{i} \gamma^{a} \phi_{a}{}^{i} - \frac{1}{2} X^{-1} \bar{\Omega}_{i} \slashed{D} \Omega^i\nn \\
	&& + \frac{1}{8} \bar{\psi}_{a} \gamma^a \slashed{D} \Omega_{i} + h.c. - \frac{1}{32} \bar{\Omega}_{i} \varepsilon^{ij} \bar{X}^{-1} \gamma \cdot (\mathcal{F}^{-} + i \mathcal{G}^{-} ) \gamma^{a} \psi_{a j}.
	\end{eqnarray}
	Apart from the dilaton Weyl multiplet, we will discuss some other multiplets in four dimensional $N=2$ conformal supergravity that are useful for our purpose. They are Tensor multiplet, Vector multiplet, Chiral multiplet and Yang-Mills multiplet.
	\subsection{Tensor Multiplet }
	The components of a tensor multiplet along with their weights and chirality are tabulated in the table below.
	\begin{table}[H]
		\caption{N=2 superconformal tensor multiplet.}\label{Table-tensor}
		\begin{center}
			\begin{tabular}{ | p{1cm}|p{2cm}|p{1cm}|p{1cm}|p{2cm}| }
				\hline
				Field & SU(2) Irreps & Weyl weight (w) & Chiral weight (c) & Chirality fermions \\ \hline
				$L_{ij}$ & $\bf{3}$ & 2 & 0 & -- \\ \hline
				$E_{\mu\nu}$ & $\bf{1}$ &  0 & 0 & -- \\ \hline
				$G$ & $\bf{1}$ & 3 & 1 & -- \\ \hline
				$\phi_{i}$ & $\bf{2}$ & 5/2 & -1/2 & -1 \\ \hline
			\end{tabular}
		\end{center}
	\end{table}
	$L_{ij}$ is an SU(2) triplet of scalar field that satisfies the reality condition $L^{ij}=\varepsilon^{ik}\varepsilon^{jl}L_{kl}$\footnote{We are following the chiral notation where complex conjugation is denoted by raising/lowering the SU(2) index. In this notation $L^{ij}=(L_{ij})^{*}$}. $E_{\mu\nu}$ is a tensor gauge field which comes with its own tensor gauge symmetry $\delta E_{\mu\nu}=2\partial_{[\mu}\lambda_{\nu]}$. $G$ is a complex scalar field and $\phi_{i}$ is an SU(2) doublet of Majorana spinors with chirality -1\footnote{In the chiral notation $\phi^{i}$ is related to the conjugate of $\phi_{i}$ via the Charge conjugation matrix $C$ as $\phi^{i}=-C^{-1}\overline{\left(\phi_{i}\right)}^{T}$, (where $\overline{\phi_i}$ is the Dirac conjugate of $\phi_{i}$) and has the opposite chiral weight and chirality}. The supersymmetry transformation of the tensor multiplet fields are given as shown below \cite{Claus:1997fk, deWit:2006gn,deWit:1982na,deWit:1980lyi}:
	\begin{eqnarray}
	\delta L_{ij} &=& 2 \bar{\epsilon}_{(i} \phi_{j)} + 2 \varepsilon_{ik} \varepsilon_{jl} \bar{\epsilon}^{(k} \phi^{l)}\ ,
	\nonumber \\
	\delta \phi^{i} &=& \slashed{D} L^{ij}\epsilon_{j} +  \varepsilon^{ij} \slashed{E}\; \epsilon_{j} - G \epsilon^{i} + 2L^{ij} \eta_{j}, \nonumber \\ 
	\delta G &=& -2 \bar{\epsilon}_{i} \slashed{D}\phi^i - \bar{\epsilon}_{i}( 2L^{ij} \chi_{j} - \frac{1}{8}\gamma .T^{+} \; \phi_{j}\; \varepsilon^{ij} \varepsilon^{kl} ) + 2 \bar{\eta}_{i}\; \phi^{i}, \nonumber \\ 
	\delta E_{\mu \nu} &=& i\bar{\epsilon}^{i} \gamma_{\mu \nu} \phi^{j} \varepsilon_{ij} + 2\;iL_{ij} \varepsilon^{jk} \bar{\epsilon}^{i} \gamma_{[\mu}\psi_{\nu]k} + h.c.\;,
	\end{eqnarray}
	where the fully supercovariant derivative acting on various fields of the tensor multiplet is given as:
	\begin{eqnarray}
	D_{\mu} L_{ij} &=&  (\partial_{\mu} - 2 b_{\mu} )\;L_{ij} - \mathcal{V}_{\mu}\;^{k}\;_{(i}L_{j)k} - \bar{\psi}_{\mu(i}\phi_{j)} - \varepsilon_{ik} \varepsilon_{jl} \bar{\psi}_{\mu}^{(k} \phi^{l)}\ ,
	\nonumber \\
	D_{\mu}\phi^{i} &=& (\partial_{\mu} -\frac{1}{4}\omega_{\mu}\;^{ab} \gamma_{ab} - \frac{i}{2}A_{\mu} - \frac{5}{2} b_{\mu} )\;\phi^{i} + \frac{1}{2} \mathcal{V}_{\mu}\;^{i}\;_{j}\phi^  {j} - \frac{1}{2}(\slashed{D} L^{ij} +  \varepsilon^{ij} \slashed{E}\;) \psi_{\mu j} \nn \\ && + \frac{1}{2} G \psi_{\mu}\;^{i} - L^{ij} \phi_{\mu j}\ ,
	\nonumber \\
	D_{\mu}G &=& (\partial_{\mu} + i A_{\mu} - 3 b_{\mu} )\;G+  \bar{\psi}_{\mu i} \slashed{D}\phi^{i} +  \bar{\psi}_{\mu i}( 3 L^{ij} \chi_{j} + \frac{1}{8}\gamma .T^{+} _{jk}\; \phi_{l}\; \varepsilon^{ij} \varepsilon^{kl} ) - \bar{\phi}_{\mu i}\; \phi^{i}.\
	\nonumber \\
	\end{eqnarray}
	$E^a$ appearing in the above transformation rules is given as the dual of the 3-form field strength $E^{a}=\frac{1}{6}\varepsilon^{abcd}\mathcal{E}_{bcd}$, where the 3-form field strength of the tensor gauge field is given as:
	\begin{eqnarray}
	\mathcal{E}_{\mu\nu\rho}&=3\partial_{[\mu}E_{\nu\rho]}-\left(\frac{3i}{2}\bar{\psi}^{i}_{[\mu}\gamma_{\nu\rho]}\phi^{j}\varepsilon_{ij}+\frac{3i}{2}L_{ij}\varepsilon^{jk}\bar{\psi}^{i}_{[\mu}\gamma_{\nu}\psi_{\rho]k}+\text{h.c}\right)
	\end{eqnarray}
	\subsection{Vector Multiplet }
	The components of an $N=2$ vector multiplet is tabulated below:
	\begin{table}[H]
		\caption{N=2 superconformal vector multiplet.}\label{Table-vector}
		\begin{center}
			\begin{tabular}{ | p{1cm}|p{2cm}|p{1cm}|p{1cm}|p{2cm}| }
				\hline
				Field & SU(2) Irreps & Weyl weight (w) & Chiral weight (c) & Chirality fermions \\ \hline
				$A$ & $\bf{1}$ & 1 & -1 & -- \\ \hline
				$C_{\mu}$ & $\bf{1}$ & 0 & 0 & -- \\ \hline
				$Y_{ij}$ & $\bf{3}$ & 2 & 0 & -- \\ \hline
				$\lambda_{i}$ & $\bf{2}$ & 3/2 & -1/2 & +1 \\ \hline
			\end{tabular}
		\end{center}
	\end{table}
	$Y_{ij}$ satisfies the reality condition $Y^{ij}=\varepsilon^{ik}\varepsilon^{jl}Y_{kl}$. $C_{\mu}$ is a vector gauge field that comes with its own vector gauge transformation $\delta C_{\mu}=\partial_{\mu}\lambda$. The supersymmetry transformations of the components are given as \cite{Claus:1997fk,deWit:2006gn,Mohaupt:2000mj,deWit:1980lyi}:
	\begin{eqnarray}
	\delta A &=& \bar{\epsilon}^{i} \lambda_{i}   \ ,
	\nonumber \\
	\delta \lambda_{i} &=& 2 \slashed{D} A \epsilon_{i} +\frac{1}{2} \gamma . \mathcal{\hat{B}}^{ -} \varepsilon_{ij} \epsilon^{j} + Y_{ij} \epsilon^{j} + 2 A \eta_{i}    \ ,
	\nn \\
	\delta C_{\mu} &=&  \varepsilon^{ij} \bar{\epsilon}_{i} \gamma_{\mu} \lambda_{j} + 2\;\varepsilon^{ij} \bar{\epsilon}_{i} \psi_{\mu j} A + h.c.  \ ,
	\nonumber\\
	\delta Y_{ij} &=& 2\; \bar{\epsilon}_{(i}  \Pi_{j)} + 2 \;\varepsilon_{ik} \varepsilon_{jl} \bar{\epsilon}^{(k} \Pi^{l)} 
	\end{eqnarray}
	with 
	\begin{align}
	\Pi_{j} = \slashed{D} \lambda_{j} - 2 A \chi_{j} 
	\end{align}
	The field strength $\mathcal{\hat{B}}_{\mu \nu} $ that appears in the above transformation laws is the modified superconformal field strength for the gauge field $C_{\mu}$ and is given as:
	\begin{equation}
	\mathcal{\hat{B}}_{\mu \nu}  = \mathcal{B}_{\mu \nu} - \frac{1}{4} \bar{A}\;T_{\mu \nu}^{-}- \frac{1}{4} {A}\;T_{\mu \nu}^{+}
	\end{equation}
	where the fully superconformal field strength $\mathcal{B}_{\mu \nu}$ is given as: 
	\begin{eqnarray}
	\mathcal{B}_{\mu \nu}  &=& 2 \partial_{[\mu} C_{\nu]} - \bar{\psi}_{[\mu \;i}\gamma_{\nu]} \lambda_{j}\varepsilon^{ij} - \bar{\psi}_{[\mu}{}^{i}\gamma_{\nu]} \lambda^{j}\varepsilon_{ij}  \ 
	\nonumber \\
	&& -A \bar{\psi}_{\mu i} \psi_{\nu j} \varepsilon^{ij} - \bar{A}\; \bar{\psi}_{\mu}{}^{ i} \psi_{\nu}{}^{ j} \varepsilon_{ij}  ,
	\end{eqnarray}
	The fully supercovariant derivative acting on various fields are given as:
	\begin{eqnarray}
	D_{\mu} Y_{ij} &=&  (\partial_{\mu} - 2 b_{\mu} )\;Y_{ij} - \mathcal{V}_{\mu}\;^{k}\;_{(i}Y_{j)k} - \bar{\psi}_{\mu(i} \slashed{D}\lambda_{j)} - \varepsilon_{ik} \varepsilon_{jl} \bar{\psi}_{\mu}^{(k}\slashed{D}\lambda^{l)}\ ,
	\nonumber \\
	D_{\mu}\lambda_{i} &=&  (\partial_{\mu} -\frac{1}{4}\omega_{\mu}\;^{ab} \gamma_{ab} + \frac{i}{2}A_{\mu} - \frac{3}{2} b_{\mu} )\;\lambda_{i} - \frac{1}{2}\mathcal{V}_{\mu}\;^{j}\;_{i}\lambda _{j} \nonumber\\
	&& -\slashed{D} A \psi_{\mu \;i} -  \frac{\varepsilon_{ij}}{4}\gamma.\hat{\mathcal{B}}\; \psi_{\mu}\;^{i} -  \frac{1}{2} Y_{ij} \psi_{\mu}^{ j} - A \phi_{\mu i} \ ,
	\nonumber \\
	D_{\mu}A &=& (\partial_{\mu} - b_{\mu} + iA_{\mu} ) A -  \frac{1}{2}\bar{\psi}_{\mu }^{i} \lambda_{i}  
	\end{eqnarray}
	\subsection{Chiral Multiplet }
	A chiral multiplet has 16 + 16 off-shell components and is a reducible multiplet. The components of a chiral multiplet along with their Weyl weights\footnote{To use chiral density formula (\ref{cdf}), $\hat{C}$ should have $w_C = 4,c_C = 0$ such that the action $S = \int d^4{}x \cL$ is  Weyl and chirally invariant. Thus,  we choose the lowest component to have weights $w_{\hat{A}} = -c_{\hat{A}} = 2$}, chiral weights, and chirality are tabulated in the table below \cite{Mohaupt:2000mj,deRoo:1980mm,deWit:1980lyi}.
	\begin{table}[H]
		\caption{N=2 superconformal chiral multiplet.}\label{Table-Chiral Multiplet}
		\begin{center}
			\begin{tabular}{ | p{1cm}|p{2cm}|p{1cm}|p{1cm}|p{2cm}| }
				\hline
				Field & SU(2) Irreps & Weyl weight (w) & Chiral weight (c) & Chirality fermions \\ \hline
				$\widehat{A}$ & $\bf{1}$ & 2 & -2 & -- \\ \hline
				$\widehat{\psi}_{i}$ & $\bf{2}$ & 5/2 & -3/2 & +1 \\ \hline
				$\widehat{B}_{ij}$ & $\bf{3}$ & 3 & -1 & -- \\ \hline
				$\widehat{F}^-_{ab}$ & $\bf{1}$ & 3 & -1 & -- \\ \hline
				$\widehat{\Lambda}_{i}$ & $\bf{2}$ & 7/2 & -1/2 & +1 \\ \hline
				$\widehat{C}$ & $\bf{1}$ & 4 & 0 & -- \\ \hline
			\end{tabular}
		\end{center}
	\end{table}
	The components are: two complex scalars $\hat{A}$ and $\hat{C}$, two $SU(2)_R$ doublets of Majorana spinors $\hat{\psi}_i$ and $\hat{\Lambda}_i$, an $SU(2)_R$ triplet of complex scalars $\hat{B}_{ij}$ (symmetric in i,j) and an antiselfdual Lorentz tensor $\hat{F}^-_{ab}$.  The supersymmetry transformations are given as follows \cite{deWit:1980lyi, deRoo:1980mm}:
	\begin{eqnarray}\label{cmt}
	\delta_{Q}(\epsilon)\hat{A} &=&\bar{\epsilon}^{i} \hat{\Psi}_{i} \nn \\
	\delta_{Q}(\epsilon) \hat{\Psi}_{i} &=&2 \slashed{D} \hat{A} \epsilon_{i}+ \hat{B}_{i j} \epsilon^{j}+\frac{1}{2}\gamma \cdot \hat{F}^{-} \varepsilon_{i j} \epsilon^{j} \nn \\
	\delta_{Q}(\epsilon) \hat{B}_{i j} &=& 2 \bar{\epsilon}_{(i} \slashed{D} \hat{\Psi}_{j)} - 2\bar{\epsilon}^{k} \hat{\Lambda}_{(i}\varepsilon_{j)k} \nn \\
	\delta_{Q}(\epsilon) \hat{F}_{a b}^{-} &=&\frac{1}{2}\varepsilon^{i j} \bar{\epsilon}_{i} \slashed{D} \gamma_{a b} \hat{\Psi}_{j}+ \frac{1}{2}\bar{\epsilon}^{i} \gamma_{a b} \hat{\Lambda}_{i} \nn \\
	\delta_{Q}(\epsilon) \hat{\Lambda}_{i} &=&-\frac{1}{2}\gamma \cdot \hat{F}^{-}\slashed{\overleftarrow{D}}  \epsilon_{i}+ \varepsilon^{k j} \slashed{D} \hat{B}_{i j} \epsilon_{k}+ \hat{C} \varepsilon_{i j}\epsilon^{j} \nn \\ &+& \frac{1}{4} \Big( \slashed{D} \hat{A} \gamma \cdot (T^+)_{ij} \varepsilon^{jk} +  \hat{A} \slashed{D} \gamma \cdot (T^+)_{ij} \varepsilon^{jk}\Big) \epsilon_{k} - 3 \gamma_{a} \varepsilon^{jk}\epsilon_{k} \underline{\bar{\chi}}_{[i}\gamma^a \hat{\Psi}_{j]} \nn\\
	\delta_{Q}(\epsilon) \hat{C} &=&-2 \varepsilon^{i j} \bar{\epsilon}_{i} \slashed{D} \hat{\Lambda}_{j} - 6 \epsilon_{i}   \chi_{j}\varepsilon^{ik}\varepsilon^{jl} \hat{B}_{kl} - \frac{1}{4} \varepsilon^{ij}\varepsilon^{kl} \epsilon_{i} \gamma \cdot (T^+)_{jk} \slashed{D}\hat{\Psi}_l 
	\end{eqnarray}
	\subsection{Yang- Mills Multiplet }
	The off-shell non-abelian $D = 4$,  $N = 2$ Yang-Mills multiplet consists of $8n + 8n$ bosonic and fermionic degrees of freedom (where n denotes the dimension of the gauge group). 
	The set of fields of the Yang-Mills multiplet along with their Weyl weights, chiral weights, and chirality are tabulated in the table below \cite{deWit:1980lyi}.
	\begin{table}[H]
		\caption{N=2 superconformal Yang-Mills multiplet.}\label{Table-Chiral Multiplet}
		\begin{center}
			\begin{tabular}{ | p{1cm}|p{2cm}|p{1cm}|p{1cm}|p{2cm}| }
				\hline
				Field & SU(2) Irreps & Weyl weight (w) & Chiral weight (c) & Chirality fermions \\ \hline
				$A^\a$ & $\bf{1}$ & 1 & -1 & -- \\ \hline
				$C_\m{}^{\a}$ & $\bf{2}$ & 0 & 0 & -- \\ \hline
				${Y}_{ij}^\a$ & $\bf{3}$ & 2 & 0 & -- \\ \hline
				${\lambda}_{i}^\a$ & $\bf{2}$ & 3/2 & -1/2 & +1 \\ \hline
			\end{tabular}
		\end{center}
	\end{table}
	$Y_{ij}^\a$ is an SU(2) triplet of scalar, $C_{\m}{}^\a$ is vector field, $A^\a$ is a complex scalar and the fermionic fields are SU(2) doublet of Majorana spinors $\lambda_i^\a$, where $\a$ denotes the Yang-Mills index ( $\a = 1.....n$).
	\noindent
	The Q- and S-transformations of the Yang-Mills multiplet are given by\footnote{In the transformation rules given in \cite{deWit:1980lyi}, there is a coupling constant $g$ that appears. We have absorbed the coupling constants by scaling the fields as $\Phi\to\frac{\Phi}{g}$}
	\begin{eqnarray*}
		\delta A^{\a} &=& \bar{\epsilon}^{i} \lambda_{i}^{\a}   \ ,
		\nonumber \\
		\delta \lambda_{i}^{\a} &=& 2 \slashed{D} A^{\a} \epsilon_{i} +\frac{1}{2} \gamma . \mathcal{\hat{B}}^{ \a-} \varepsilon_{ij} \epsilon^{j} + Y_{ij}^{\a} \epsilon^{j} + 2 A^{\a} \eta_{i}  + 2 A^{\b} \bar{A}^{\g} f_{\b \g}{}^{\a}  \varepsilon_{ij} \epsilon^{j} \ ,
		\nonumber\\
		\delta C_{\mu}^{\a} &=&  \varepsilon^{ij} \bar{\epsilon}_{i} \gamma_{\mu} \lambda_{j}^{\a} + 2\;\varepsilon^{ij} \bar{\epsilon}_{i} \psi_{\mu j} A^{\a} + h.c.  \ ,
		\nonumber\\
		\delta Y_{ij}^{\a} &=& 2\; \bar{\epsilon}_{(i}  \Pi_{j)}^{\a} + 2 \;\varepsilon_{ik} \varepsilon_{jl} \bar{\epsilon}^{(k} \Pi^{l)\a}+ 4 \varepsilon_{k(i} (\; \bar{\epsilon}_{j)} A^{\b} \lambda^{k \g} - \bar{\epsilon}^{k} \bar{A}^{\b} \lambda_{j)}{}^{\g}\;) f_{\b \g}{}^{\a} \ ,
	\end{eqnarray*}
	with 
	\begin{align}
	\Pi_{j}^\a = \slashed{D} \lambda_{j}^\a - 2 A^\a \chi_{j} 
	\end{align}
	Similar to the vector multiplet, the modified superconformal field strength $\mathcal{\hat{B}}_{\mu \nu} {}^\a$ is defined as:
	
	\begin{eqnarray}
	\mathcal{\hat{B}}_{\mu \nu}{}^{\a } &=& 2 \partial_{[\mu} C_{\n]}{}^\a - \bar{\psi}_{[\mu \;i}\gamma_{\nu]} \lambda_{j}^\a \varepsilon^{ij} - \bar{\psi}_{[\mu}\;^{i}\gamma_{\nu]} \lambda^{j \a} \varepsilon_{ij} +  C_{\n}{}^{\b } C_{\m}{}^{\g} f_{\b \g}{}^\a \ 
	\nonumber \\
	&& -A^\a \bar{\psi}_{\mu i} \psi_{\nu j} \varepsilon^{ij} - \bar{A}^\a\; \bar{\psi}_{\mu}\;^{ i} \psi_{\nu}\;^{ j} \varepsilon_{ij}- \frac{1}{4} \bar{A}^\a\;T_{\mu \nu}{}^{-} - \frac{1}{4} {A}^\a\;T_{\mu \nu}{}^{+} ,
	\end{eqnarray}
	The fully supercovariant derivative acting on various fields are given as:
	\begin{eqnarray}
	D_{\mu} Y_{ij}^\a &=&  (\partial_{\mu} - 2 b_{\mu} )\;Y_{ij}^\a - \mathcal{V}_{\mu}\;^{k}\;_{(i}Y_{j)k}^\a - \bar{\psi}_{\mu(i} \slashed{D}\lambda_{j)}^\a - \varepsilon_{ik} \varepsilon_{jm} \bar{\psi}_{\mu}^{(k}\slashed{D}\lambda^{m \a)}\ ,
	\nonumber \\
	D_{\mu}\lambda_{i}^\a &=&  (\partial_{\mu} -\frac{1}{4}\omega_{\mu}\;^{ab} \gamma_{ab} + \frac{i}{2}A_{\mu} - \frac{3}{2} b_{\mu} )\;\lambda_{i}^\a - \frac{1}{2}\mathcal{V}_{\mu}\;^{j}\;_{i}\lambda _{j}^\a \nonumber\\
	&& -\slashed{D} A^\a \psi_{\mu \;i} -  \frac{1}{4} \varepsilon_{ij} \gamma.\hat{\mathcal{B}}^\a\; \psi_{\mu}^{j} -  \frac{1}{2} Y_{ij}^\a \psi_{\mu}^{ j} - A \phi_{\mu i} + \; C_{\mu} ^{\b} \lambda_{i}^{\g} f_{\b \g}{}^{\a} \ ,
	\nonumber \\
	D_{\mu}A^\a &=& (\partial_{\mu} - b_{\mu} + iA_{\mu} ) A^\a -  \frac{1}{2}\bar{\psi}_{\mu }^{i} \lambda_{i}^\a   + \; C_{\mu} ^{\b} A^{\g} f_{\b \g}{}^{\a} 
	\end{eqnarray}
	\section{Superconformal Actions}{\label{3}}
	Superconformally invariant actions in conformal supergravity are obtained by exploiting the knowledge of density formulae. Density formulae in conformal supergravity are superconformally invariant actions given either in terms of an abstract multiplet or in terms of a known multiplet. In order to obtain an invariant action for a given multiplet, all we need to do is to embed the given multiplet into the multiplet in terms of which the density formula is given. In this section, we will discuss the relevant density formulae that we use to construct superconformal actions for the various multiplets discussed in the previous section. Using these density formulae we can construct superconformal action for tensor multiplet \cite{deWit:1982na} as well as Yang-Mills multiplet \cite{deWit:1980lyi}. In order to do that, we need to embed the tensor multiplet and the Yang-Mills multiplet into a vector multiplet and chiral multiplet respectively. We will discuss this in subsections-\ref{tcf} and \ref{ycf}.
	\subsection{Density Formulae}
	In this subsection, we will discuss two density formulae: one given in terms of a tensor multiplet and a vector multiplet, also known as tensor-vector density formula (\ref{df}), and the second one is given in terms of a chiral multiplet, also known as the chiral density formula (\ref{cdf}). The former is used for the construction of a superconformally invariant action for the tensor multiplet action and we use the latter to construct the superconformally invariant action of a Yang-Mills multiplet.
	
	\noindent 
	The tensor-vector density formula is given as \cite{deWit:1982na,Claus:1997fk,deWit:2006gn}: 
	\begin{eqnarray}\label{df}
	e^{-1} \mathcal{L} &=& AG -  \frac{1}{4} Y^{ij} L_{ij} - \frac{i}{8} \varepsilon^{abcd}   E_{ab} \cB_{cd} + \bar{\phi}^{i} ( \lambda_{i} + A \;\gamma^{a} \psi_{a \;i} )\nonumber\\
	&&  - \frac{1}{2} (\bar{\psi}_{a}^{i} \gamma^{a} \lambda^{j} + \bar{A}\; \bar{\psi}_{a} ^{i} \gamma^{ab} \psi_{b} ^{j} ) L_{ij}+\text{h.c}
	\end{eqnarray}
	Superconformally invariant chiral density formula is given as follows \cite{deRoo:1980mm,Mohaupt:2000mj}:
	\begin{eqnarray}\label{cdf}
	e^{-1} \mathcal{L}&=& \hat{C}-\bar{\psi}_{i} \cdot \gamma \Lambda_{j} \varepsilon^{i j}+\frac{1}{8} \bar{\psi}_{a i} \gamma \cdot T^{+} \gamma^{a} \Psi_{j} \varepsilon^{i j}-\frac{1}{4} A T_{a b}^{+} T^{+a b}-\frac{1}{2} \bar{\psi}_{a i} \gamma^{ab} \psi_{b j} B_{k l} \varepsilon^{i k} \varepsilon^{j l} \nn\\
	&&+\bar{\psi}_{a i} \psi_{b j} \varepsilon^{i j} G^{-ab}-\bar{\psi}_{a i} \psi_{b j} \varepsilon^{i j} A T^{+ab}+\frac{1}{2}  \varepsilon^{i j} \varepsilon^{k l}  \varepsilon^{abcd} \bar{\psi}_{a i} \psi_{b j} \bar{\psi}_{c k} \gamma_{d} \Psi_{l}\nn \\
	&&+\frac{1}{2}  \varepsilon^{i j} \varepsilon^{k l}  \varepsilon^{abcd} \bar{\psi}_{a i} \psi_{b j} \bar{\psi}_{c k} \psi_{d l} A+\mathrm{h.c.}
	\end{eqnarray}
	
	\subsection{Tensor Multiplet Action}\label{tcf}
	We can identify the following composite of the tensor multiplet fields that transform as the vector multiplet scalar: 
	\begin{align}
	B = G^{*} L^{-1} - \bar{\phi}_{i} \phi_{j} L^{ij} L^{-3}
	\end{align}  
	where $ L^{2} = {L_{ij}L^{ij}}$. 
	This identification has the correct chiral and Weyl weight and is invariant under S-supersymmetry. Upon successive application of supersymmetry transformation one obtains the complete expressions for the components of a vector multiplet in terms  of the components of tensor multiplet. The embedding are as follows \cite{deWit:1982na}
	\begin{eqnarray} \label{m}
	B &=& G^{*} L^{-1} - \bar{\phi}_{i} \phi_{j} L^{ij} L^{-3} , \nonumber \\
	\Sigma _{i} &=&   - 2 L^{-1}\;\slashed{D}\;\phi_{i} -(2 L_{ij}\chi^{j} - \frac{1}{4} \varepsilon_{il}\; \phi^{l} \;\gamma\;. T^{-} )\; L^{-1} \nonumber \\
	&& - 2 L_{ij} \phi^{j} G^{*} L^{-3} + 2( \slashed{D} L_{ij} L^{jk} \;\phi_{k} - \varepsilon_{ij} L^{jk} \phi_{k} \slashed{E} ) L^{-3} \nonumber \\
	&& -2 \bar{\phi}_{i}\phi_{k} \phi^{k} L^{-3} + 6 \phi^{j}\bar{\phi}_{k}\phi_{m} L^{km} L_{ij} L^{-5} ,\nonumber\\
	\mathcal{Z}_{ij} &=&  - L^{-1} \square^{c} L_{ij} - DL^{-1} L_{ij} + D_{a}L_{ik}D^{a}L_{jl} L^{kl} L^{-3} + E^{2}L_{ij} L^{-3} \nonumber\\
	&&+ |G|^{2} L_{ij} L^{-3} -2 E_{\mu} D^{\mu}L_{k(i} \varepsilon_{j)l}L^{kl} L^{-3} + 2 G L^{-3} \bar{\phi}_{i} \phi_{j} \nonumber\\
	&& - 6 G L_{ij} \bar{\phi}_{k} \phi_{l} L^{kl} L^{-5} - \varepsilon_{k(i} \bar{\phi}^{k} \slashed{E} \phi_{j)} L^{-3} -4 L_{k(i}\bar{\phi}^{k} \slashed{D}\phi_{j)} L^{-3}   \nonumber\\
	&&  - 6L_{k(i}L_{j)l} \bar{\phi}^{l} \slashed{E} \phi_{m}\varepsilon^{km} L^{-5}- \bar{\phi}^{k} \slashed{D} L_{k(i}\phi_{j)} L^{-3} - \bar{\phi}^{k} \slashed{D}L_{ij} \phi_{k} L^{-3}\nonumber\\
	&& +\; 6L_{k(i}\bar{\phi}^{k} \slashed{D}L_{j)l}\phi_{m}L^{lm} L^{-5} - 6 L_{k(i}\bar{\phi}^{k} \phi^{l} \bar{\phi}_{j)} \phi_{l} L^{-5} + \frac{3}{2} \bar{\phi}_{i} \phi_{j} \bar{\phi}^{k} \phi^{l} L_{kl} L^{-5} \nonumber\\
	&& + 15 L_{ik} L_{jl} \bar{\phi}^{k} \phi^{l} \bar{\phi}_{m}\phi_{n} L^{mn} L^{-7}   - \frac{1}{2} \varepsilon_{k(i}L_{j)l}\bar{\phi}^{l}\gamma . T^{-}  \phi^{k} L^{-3} \nonumber\\
	&& - 12 L_{k(i}L_{j)l} \bar{\chi}^{k} \phi^{l} L^{-3} + 2\bar{\chi}_{(i}\phi_{j)}L^{-1}  + \varepsilon_{ik}\varepsilon_{jl} [h.c.]^{kl} \, \nonumber\\
	{\mathcal{C}}_{ab} &=& 2 D_{[a} L_{ik} D_{b]}L_{jl}L^{kl} \varepsilon^{ij} L^{-3} - 4 D_{[a} {\left( E_{b]} L^{-1}\right)}  -\varepsilon^{ij} L^{-1} L_{ik} R_{ab}(V)^{k}\;_{j}  + fermions \nonumber\\
	\end{eqnarray}
	
	\noindent
	Using the tensor-vector density formula (\ref{df}) and the embedding of the tensor multiplet in the vector multiplet given above (\ref{m}) we can write down the superconformal action for tensor multiplet. The bosonic part of superconformal tensor multiplet action is given as follows:
	
	\begin{eqnarray}\label{tma}
	e^{-1} \mathcal{L}_{T} &=& DL + L^{-1} L^{ij} \square^{c} L_{ij}  + |G|^{2}L^{-1}-E^2 L^{-1}+ 2i  \tilde{E}^{ab} D_{a}\left(E_{b}L^{-1}\right) \nonumber\\
	&&+\frac{i}{2}\varepsilon^{i j} L_{ki} \tilde{E}^{ab} L^{-1} R_{ab}(V)^{k}\;_{j}  -  L^{ij} D_{a}L_{ik} D^{a}L_{jl} L^{kl} L^{-3} \nonumber\\
	&& -i  \tilde{E}^{ab} D _{a} L_{ik} D _{b} L_{jl}  L^{kl} \varepsilon ^{ij} L^{- 3}
	\end{eqnarray}
	where $ \tilde{E}^{ab} $ is the dual of the tensor gauge field $E_{ab}$ given as:
	\begin{align}
	\tilde{E}^{ab} =  \frac{1}{2} \varepsilon^{abcd} E_{cd}
	\end{align}
	
	\noindent 
	The fully superconformal d'Alambertian for $L_{ij}$ is given by 
	\begin{eqnarray}
	\square^{c} L_{ij} = D^{a} D_{a} L_{ij}  &=&  \mathcal{D}^{a} D_{a} L_{ij} + 2 f_{a}^{a} L_{ij} - (\; \bar{\psi}_{\mu (i} D^{\mu} \phi_{j)} + \varepsilon_{ik} \varepsilon_{jl}\;\bar{\psi}_{\mu}^{ (k} D^{\mu} \phi^{l)}\;)  \nonumber\\
	&& - 2 \left(\; \bar{\psi}_{\mu (i} \gamma^{\mu} \chi^{k} L_{j)k}  - \bar{\psi}_{\mu}\;^{k} \gamma^{\mu} \chi_{(i} L_{j)k}  + \bar{\psi}_{\mu}\;^{k} \gamma^{\mu} \chi_{k} L_{ij} \right) \nonumber \\
	&& - \frac{1}{16} \left( \bar{\phi}_{( i} \gamma . T_{j)k}  \gamma^{\mu} \psi_{\mu}\;^{k} +  \varepsilon_{ik} \varepsilon_{jl}  \bar{\phi}^{( k} \gamma . T\;^{l)m}  \gamma^{\mu} \psi_{\mu m} \right) 
	\nonumber \\
	&& - \frac{1}{2} \left( \bar{\phi}_{\mu ( i} \gamma^{\mu} \phi_{j)} +  \varepsilon_{ik} \varepsilon_{jl} \bar{\phi}_{\mu} ^{(k} \gamma^{\mu} \phi^{l)} \right)
	\end{eqnarray}

	
	
	

	\subsection{Superconformal Yang-Mills Action}\label{ycf}

	\noindent
	The lowest component of the chiral multiplet in a chiral density formula needs to have $w=-c=2$, where $w$ and $c$ are the Weyl and chiral weights respectively. In order to use the chiral density formula to construct an invariant action for the Yang-Mills multiplet we need to find a composite which has the required property of the lowest component of a chiral multiplet. The composite is given as follows:
	\begin{align}
	\hat{A} = \cA = A^{\a} A^\b Tr(T_\a T_\b ) 
	\end{align}
	where $A^\a$ is the  complex scalar of a Yang-Mills multiplet with Weyl weight $w=1$ and chiral weight $c=-1$.
	\noindent 
	One can see that $\hat{A}$ transforms as the lowest component of a $w = 2$ and $c=-2$ chiral multiplet and by successive application of supersymmetry transformation, we can obtain all the components of the $ w = 2$ chiral multiplet. 
	\begin{eqnarray}\label{ycf1}
	\hat{A} &=& \cA = A^{\a} A^\b a_{\a\b} \nn \\
	\hat{\psi}_i &=& 2A^\a \lambda_i^\b a_{\a \b} \nn \\
	\hat{B}_{ij} &=& 2 A^\a Y_{ij} ^\b a_{\a\b} - \bar{\lambda}_i^\a \lambda_j^\b a_{\a \b} \nn \\
	\hat{F}^{-}_{ab} &=& 2 A^{\a} \hat{B}^{-\b}_{ab} a_{\a\b} - \frac{1}{4} \varepsilon^{kl} \bar{\lambda}_k^\a \gamma_{ab} \lambda_l^\b a_{\a\b} \nn \\ 
	2 \hat{\Lambda} &=& - \Big( \gamma \cdot \hat{\cB}^{-\a} \Big) \lambda_i ^{\b} a_{\a \b} + 2 \varepsilon^{lm} \lambda_l ^\a Y_{mi}^\b a_{\a \b} + 4 \varepsilon_{mi} A^\a \Pi^{m \b} a_{\a \b} - 4 A^\g \bar{A}^\d \lambda_i^\b f_{\g \d}{}^\a a_{\a \b} \nn \\
	2 \hat{C} &=& 2 \hat{\cB}^{-\a} \cdot \hat{\cB}^{-\b}  a_{\a \b}  - 8 A^\a \square^c \bar{A}^\b a_{\a \b} + 8 A^\a \bar{A}^\b D a_{\a \b} - A^\a \hat{\cB}^{\b} \cdot T^+ a_{\a \b} \nn \\
	&&- Y_{ij}^\a Y^{\b ij} a_{\a\b} - 8 [ A , \bar{A} ]^2 + fermions 
	\end{eqnarray}
	where, we have defined $a_{\a\b}\equiv Tr(T_\a T_\b)$. Using the chiral density formula (\ref{cdf}) and the embedding given in (\ref{ycf1} ) one obtains the superconformal action for Yang-Mills multiplet. The bosonic part of  the action is given as follows:
	\begin{eqnarray}\label{yma}
	\Big(e^{-1} \mathcal{L}_{YM}\Big)^{bosonic} &=&  8 \mathcal{D'}_a A^\a \mathcal{D'}^a \bar{A}^\b a_{\a \b } + 8i A_a  A^\a \overleftrightarrow{\cD'}^a \bar{A}^\b  a_{\a \b} + 8 A_a A^a A^\a \bar{A}^\b a_{\a \b}  \nn \\ &-&  8 f_{a}{}^a A^\a \bar{A}^\b a_{\a \b}  + 8 D A^\a \bar{A}^\b a_{\a \b} -  Y_{ij}^\a Y^{\b ij } a_{\a \b}  -  8 [ A, \bar{A}]^2 \nonumber\\ &+& 2\hat{\cB}^{- \a} \cdot \hat{\cB}^{-\b }a_{\a \b} -  A^\a \hat{\cB}^{\b} \cdot T^{+} a_{\a \b}  + \text{h.c}  
	\end{eqnarray}
	where the D is the composite field given in expression (\ref{T}) 
	and 
	\begin{align*}
	\mathcal{D'}_\m A^\a = ( \partial_\m - b_\m )A^\a +  C_{\m}{}^\b A^\g f_{\b \g}{}^\a 
	\end{align*}
	Now that we have obtained the superconformal actions, we are in a position to write the actions in super-Poincar\'e background by breaking the superfluous symmetries in the superconformal theory. In the next section we would discuss the gauge fixing of superconformal group to super-Poincar\'e group and subsequently we will write down the actions in super-Poincar\'e theory  using the decomposition rules.  
	
	\section{Gauge fixing and super-Poincar\'e transformation}\label{4}
	We can break special conformal symmetry by simply setting the dilatation gauge field to zero:
	\be
	K-gauge: \;\;\;\;b_{\mu} = 0 
	\ee
	\noindent
	But the gauge condition itself is not invariant under dilatation, Q-supersymmetry as well as S-supersymmetry. Hence, we need to modify the transformation laws under Q-supersymmetry, S-supersymmetry, and dilatation by a compensating field dependent K- transformation which is given by the following expression:
	\begin{align}\label{K-comp}
	\Lambda_{K \m} = - \frac{1}{2} \bar{\epsilon}^i \phi_{\mu i} + \frac{1}{2} \bar{\epsilon}^{i} \gamma_{\mu} \chi_i + \frac{1}{2} \bar{\eta}^{i} \psi_{\mu i} + \text{h.c}-\partial_{\mu}\Lambda_{D}
	\end{align}
	\noindent
	The $U(1)_{R}$ symmetry can be broken by imposing the following condition on the dilaton Weyl multiplet scalar $X$:
	\be 
	A-gauge:\;\;\;\;X = \bar{X}
	\ee
	and to break dilatation, we make the following choice:
	\be
	D-gauge:\;\;\;\;X = \bar{X} = 1
	\ee
	In order to break S-supersymmetry, we choose the following gauge condition,
	\be
	S-gauge:\;\;\;\Omega_i = 0
	\ee 
	The S-gauge condition breaks Q-supersymmetry and hence needs to be compensated by field dependent S-supersymmetry parameter which is given as shown below:
	\begin{align}\label{S-comp}
	\eta^i &= - \frac{1}{12} \gamma^{a}\varepsilon_{abcd} \cH^{bcd} \epsilon^{i} - \frac{1}{8}\varepsilon^{ij} \gamma \cdot \mathcal{F} \epsilon_{j} - \frac{i}{8} \varepsilon^{ij} \gamma \cdot \mathcal{G}^{+} \epsilon_{j}
	\end{align} 
	\noindent
	A-gauge, D-gauge, and S-gauge together implies that we do not need any compensating parameter for dilatation and U(1) R-symmetry. By taking into account the compensating parameters for special conformal transformation and S-supersymmetry, the Poincar{\'e} supersymmetry transformation is given as:
	\be \label{4.7}
	\delta^{P}(\epsilon) = \delta_{Q}(\epsilon) + \delta_{S}(\eta) + \delta_{K}(\Lambda_{K})
	\ee
	where on the R.H.S the compensating parameters $\eta$ and $\Lambda_{K}$ are functions of $\epsilon$ via equations (\ref{S-comp}) and (\ref{K-comp}). Using the above prescription, we obtain the Poincar{\'e} supersymmetry transformations of the dilaton Weyl multiplet as:
	\begin{eqnarray}
	\delta^P e_{\mu} \;^{a} &=& \bar{\epsilon}^{i} \gamma^{a} \psi_{\mu i} + \text{h.c} +  \Lambda _{M}^{ab} e_{\mu b}, \nonumber \\
	\delta^P \psi_{\mu}\;^{i} &=& 2 \cD_{\m}(\omega_+)\epsilon^i - \frac{1}{2}\varepsilon^{ij} \gamma^{\r} (\mathcal{F}_{\r \m} + i \mathcal{G}_{\r \m}) \epsilon_j,\nn \\
	\delta^P W_\mu &=&  2 \varepsilon_{ij}  \bar{\epsilon}^i \psi_\m{} ^j + \text{h.c} + \partial_\m \lambda \;, \nn \\ 
	\delta^P \tilde{W}_\mu &=& - 2\;i\; \varepsilon_{ij}  \bar{\epsilon}^i \psi_\m {}^j + \text{h.c} + \partial_\m \tilde{\lambda} \;, \nn\\
	\delta^P B_{\m \n} &=& \frac{1}{2} W_{[\m}\delta_{Q} W_{\n]} + \frac{1}{2} \tilde{W}_{[\m}\delta_{Q} \tilde{W}_{\n]} +  2  \bar{\epsilon}^i \gamma_{[\m}\psi_{\n] i} + 2  \bar{\epsilon}_i \gamma_{[\m}\psi_{\n]}{}^{ i} \nn \\ 
	&& + 2 \partial_{[\m} \Lambda_{\n]} - \frac{\lambda}{4} F_{\m \n} - \frac{\tilde{\lambda}}{4} G_{\m \n}, \nn \\ 
	\delta^P {{V}}_{\mu}{}^{i}{}_{j}  &=& 2 \bar{\epsilon}_{j} \gamma^c ({\widehat{\psi}}_{c \m})^i + 4 i \bar{\epsilon}_{j} \gamma^c A_{c} \psi_{\m}{}^i - \text{(h.c.; traceless)}\\
	\end{eqnarray}
	where the torsionful spin connection $\omega_{\pm \; \m}{}^{ab}$  is defined as
	\be \label{tsc}
	\omega_{\pm \;\m}{}^{ab} = \omega_{\m}{}^{ab} \pm \frac{1}{2}\mathcal{H}_{\m}{}^{ab} 
	\ee
	It is interesting to note that the field strength $\mathcal{H}_{\m \n \r}$ has a natural torsion interpretation in this gauge choice. The Poincar{\'e} supersymmetry transformation of the spin-connection would be required later and can be obtained as:	
\begin{eqnarray} 
	\delta^P \omega_{\m ab} &=& \frac{1}{2}\bar{\epsilon}_{i} \gamma_{\m} R'_{ab}(Q)^i + \frac{i}{2} \bar{\epsilon}_i \gamma_\m A_{[a}\psi_{b]}{}^i - \bar{\epsilon}_i \gamma_{[a}R'_{b]\m}(Q)^i + \frac{i}{2} \bar{\epsilon}_{i} A_{\m} \gamma_{[a}\psi_{b]}{}^i  - \frac{i}{2} \bar{\epsilon}_i \gamma_{[a}A_{b]}\psi_\m{}^i  \nn \\ && - \frac{i}{2} \bar{\epsilon}_i \gamma_{\r a b} A^\r \psi_{\m}{}^i - i \bar{\epsilon}_{i} A_{[a}\gamma_{b]}\psi_{\m}{}^i - i \bar{\epsilon}_{i} \eta_{\m[a} \psi_{b]}{}^i\gamma_{\r}A^\r  + \frac{1}{16} \varepsilon^{ij} \bar{\epsilon}_{i} \gamma \cdot (\mathcal{F}  + i \mathcal{G}) \gamma_{ab} \psi_{\m j} \nn \\ && +\frac{1}{8} \varepsilon^{ij} \bar{ \epsilon}_{i} \gamma \cdot (\mathcal{F} + i \mathcal{G}) \eta_{\m [a} \psi_{b] j} - \frac{1}{2} \varepsilon^{ij} \bar{\epsilon}_i (\mathcal{F}_{ab}{}^- + i \mathcal{G}_{ab}{}^-)\psi_{\m j} + \text{h.c}
	\end{eqnarray}

	We can write down the fully supercovariant curvatures for the gravitino, $SU(2)$ gauge field, the two $U(1)$ gauge field $W_{\mu}$ and $\tilde{W}_{\mu}$ as well as the tensor gauge field $B_{\mu\nu}$ in the super-Poincar\' e theory as follows 
	\bea\label{sc1}
	\widehat{\psi}_{\mu\nu}{}^i &=& 2 \cD_{[\m}(\omega_{+})\psi_{\n]}{}^i - \frac{1}{2}\varepsilon^{ij} \gamma^{\rho} {K}_{\rho[\mu}\psi_{\nu]}{}_j  ,\,\nn\\
	(\widehat{V}_{\mu\nu})^i{}_j &=& 2 \partial_{[\m}\cV_{\nu]}^i{}_j + \cV^i{}_{[\m k}\cV^k{}_{\n]j} + {} [2 \bar{\psi}_{j[\m}\gamma^\r \widehat{\psi}_{\n] \r}{}^i - {2}i{}\bar{\psi}_{j[\m}\gamma^c A_c \widehat{\psi}_{\n]}{}^i - \text{(h.c.; traceless)}] \nn \\
	{\mathcal{F}}_{\m \n} &=& 2 \partial_{[\m}W_{\n]} - \varepsilon_{ij} \bar{\psi}_{[\mu}^{i} \psi_{\n]}{}^j - \varepsilon^{ij} \bar{\psi}_{[\m i}\psi_{\n] j} ,\,\nn\\
	{\mathcal{G}}_{\m \n} &=& 2 \partial_{[\m}\tilde{W}_{\n]} + i\; \varepsilon_{ij} \bar{\psi}_{[\mu}^{i} \psi_{\n]}{}^j - i\; \varepsilon^{ij} \bar{\psi}_{[\m i}\psi_{\n]j} ,\,\nn\\
	\widehat{\mathcal{H}}_{\m \n \r} &=& 3 \partial_{[\m} B_{\n \r]} + \frac{3}{4}W_{[\m}F_{\n \r]} + \frac{3}{4}\tilde{W}_{[\m}G_{\n \r]} - 3 \bar{\psi}_{[\mu}{}^i \gamma_{\n}\psi_{\r]i}.
	\eea
	where we have defined:
	\bea
	{K}_{\m\n} &=& \mathcal{F}_{\m\n} + i \mathcal{G}_{\m\n}, \, \nn\\
	{\bar{K}}_{\m\n} &=& \mathcal{F}_{\m\n} - i \mathcal{G}_{\m\n}.
	\eea
The spacetime index $\mu, \nu \cdots$ in the curvatures defined above can be converted into local Lorentz index $a,b \cdots$ by contracting them with inverse vielbein and using the constraint equation $2\cD_{[\mu}e_{\n]}^{a}=\bar{\psi}_{[\mu}^{i}\gamma^{a}\psi_{\nu i}$. For the curvature associated with supersymmetry and $SU(2)$-R symmetry, we get (similar expression can be obtained for the other curvatures as well):
\begin{align}
\widehat{\psi}_{ab}^{i}&=2 \cD_{[a}(\omega,\omega_{+})\psi_{b]}{}^i - \frac{1}{2}\varepsilon^{ij} \gamma^{c} {K}_{c[a}\psi_{b]}{}_j 	+\bar{\psi}_{[a}^{j}\gamma^{c}\psi_{b]j}\psi_{c}^{i}\nn\\
(\widehat{V}_{ab})^i{}_j &= 2 \partial_{[a}\cV_{b]}^i{}_j +2\omega_{[ab]}{}^{c}\cV_{c}{}^{i}{}_{j}+ \cV^i{}_{[a k}\cV^k{}_{b]j}  +\bar{\psi}_{[a}^{k}\gamma^{c}\psi_{b]k}\cV_{c}{}^{i}{}_{j}  \nn\\
&+ {} [2 \bar{\psi}_{j[a}\gamma^c \widehat{\psi}_{b] c}{}^i - {2}i{}\bar{\psi}_{j[a}\gamma^c A_c \widehat{\psi}_{b]}{}^i - \text{(h.c.; traceless)}]
\end{align}
	\noindent
	It is important to note that in $ \cD_{a}(\omega, \omega_{+})\psi_{b}{}^i$, the spin connection $\omega$ rotates Lorentz vector indices and spin connection $\omega_+$ rotates Lorentz spinor index.
	
	\noindent 
	The Poincar{\'e} supersymmetry transformation for $\widehat{\psi}_{ab}{}^i$ would be required later and is obtained as:
	\bea
	\delta^P \widehat{\psi}_{ab}{}^i &=& - \frac{1}{2} \gamma^{cd}{} \widehat{R}_{cdab}(\omega_{+})\epsilon^{i} + (\widehat{V}_{ab})^i\;_j \epsilon^j + \varepsilon^{ij} D_{[a}(\omega, \omega_+) {K}_{b]c} \gamma^c \epsilon_j \nn\\
	&& + \frac{1}{4}\epsilon^i {} \gamma^{cd}{}\left( \mathcal{F}_{ca}\mathcal{F}_{bd} + \mathcal{G}_{ca}\mathcal{G}_{bd} \right) +  \frac{i}{4} \epsilon^i {} \eta^{cd}{} \left( \mathcal{G}_{ca}\mathcal{F}_{bd} - \mathcal{G}_{bd}\mathcal{F}_{ca} \right)\;.
	\eea
$\widehat{R}_{\mu\nu}{}^{ab}(\omega)$ is the fully supercovariant curvature associated with $\omega_{\mu}^{ab}$ in Poincar{\'e} supergravity. At the bosonic level it coincides with the standard Riemann tensor. Upon using the Bianchi identity on supercovariant curvature $\mathcal{H}_{abc}$
	\begin{align}
	D_{[a}(\o) \mathcal{H}_{bcd]} = \frac{1}{8}{} [ \mathcal{F}_{ab}  \mathcal{F}_{cd} +  \mathcal{G}_{ab} \mathcal{G}_{cd} + 2  \mathcal{F}_{a [c} \mathcal{F}_{d]b} + 2  \mathcal{G}_{a[c} \mathcal{F}_{d]b}{}]
	\end{align}
	we obtain,
	\be \label{bi}
	\widehat{R}_{abcd}(\omega_{+}) = \widehat{R}_{cdab}(\omega_{-}) + \frac{1}{4}{} [ \mathcal{F}_{ab}  \mathcal{F}_{cd} +  \mathcal{G}_{ab} \mathcal{G}_{cd} + 2  \mathcal{F}_{a [c} \mathcal{F}_{d]b} + 2  \mathcal{G}_{a[c} \mathcal{F}_{d]b}{}]
	\ee
	Upon using equation (\ref{bi}) and Bianchi identity $D_{[a}(\o) \mathcal{F}_{bc]} = 0 = D_{[a}(\o) \mathcal{G}_{bc]}$, we obtain 
	\bea
	\delta^P \widehat{\psi}_{ab}{}^i &=& - \frac{1}{2} \gamma^{cd} \widehat{R}_{cdab}(\omega_{-})\epsilon^{i} + (\widehat{V}_{ab})^i\;_j \epsilon^j -\frac{\varepsilon^{ij}}{2} \slashed{D}(\omega_{-}) {{K}}_{ab} \epsilon_{j} - \frac{1}{4}{\mathcal{F}}_{ab} \gamma \cdot {\mathcal{F}} \epsilon^{i} \nn\\ && - \frac{1}{4}{\mathcal{G}}_{ab} \gamma \cdot {\mathcal{G}} \epsilon^{i}  + \frac{i}{4} \eta^{cd} [{\mathcal{G}}_{ac}{\mathcal{F}}_{db} + {\mathcal{F}}_{ac}{\mathcal{G}}_{db}]
	\eea
	Poincar{\'e} supersymmetry transformation of the fully supercovariant curvatures corresponding to the two $U(1)$ and the tensor gauge field  is given as\footnote{Instead of giving the supersymmetry transformation of the fully supercovariant curvature $\mathcal{H}_{abc}$, we give the transformation of $\mathcal{H}_{\mu ab}$ which will be useful for our future calculations}: 
	\bea
	\delta^P {\mathcal{F}}_{ab} &=& 2 \varepsilon_{ij}\bar{\epsilon}^{i} \widehat{\psi}_{ab}^j + \text{h.c} \,,\nn\\
	\delta^P {\mathcal{G}}_{ab} &=& - 2i \varepsilon_{ij}\bar{\epsilon}^{i} \widehat{\psi}_{ab}^j + \text{h.c} \,,\nn\\
	\delta^P {\mathcal{H}}_{\mu a b} &=& - \bar{\epsilon}_i \gamma_\m R'_{ab}(Q)^i - i \bar{\epsilon}_i \gamma_\m A_{[a }\psi_{b]}{}^i - 2 \bar{\epsilon}_{i} \gamma_{[a}R'_{b] \m}(Q)^i + i \bar{\epsilon}_{i} A_{\m} \gamma_{[a}\psi_{b]}{}^i  - i \bar{\epsilon}_i \gamma_{[a}A_{b]}\psi_\m{}^i  \nn \\ && + i{} \bar{\epsilon}_i \gamma_{\r a b} A^\r \psi_{\m}{}^i - 2 i \bar{\epsilon}_{i} A_{[a}\gamma_{b]} \psi_{\m}{}^i - 2 i \bar{\epsilon}_{i} \eta_{\m[a} \psi_{b]}{}^i\gamma_{\r}A^\r  - \frac{1}{8} \varepsilon^{ij} \bar{\epsilon}_{i} \gamma_{ab} \gamma \cdot (\mathcal{F}  + i \mathcal{G})  \psi_{\m j} \nn \\ && +\frac{1}{4} \varepsilon^{ij} \bar{\epsilon}_{i} \gamma \cdot (\mathcal{F} + i \mathcal{G}) \eta_{\m [a} \psi_{b] j} +  \varepsilon^{ij} \bar{\epsilon}_i \gamma_\m \gamma^c (\mathcal{F}_{c [a}{}^+ + i \mathcal{G}_{c[a}{}^+)\psi_{b] j} \nonumber \\
	&& + \bar{\epsilon}_i \gamma^c \mathcal{H}_{cab}\psi_{\m}{}^i + \text{h.c}
	\eea
	We can also break the $SU(2)$-R symmetry by imposing a suitable gauge condition which will further modify the Poincar{\'e} supersymmetry transformation law by a field dependent $SU(2)$-parameter. We will discuss this in Appendix-\ref{SU2breaking}. In the following sections, we will work with an unbroken $SU(2)$-R symmetry. One can implement the $SU(2)$-breaking later as per the discussions in Appendix-\ref{SU2breaking}.
	\section{$N=2$ Poincar\'e Supergravity}\label{sec5}
	In this section we will discuss construction of $N=2$ Poincar{\'e} supergravity by imposing the gauge fixing conditions on $N=2$ conformal supergravity discussed in the previous section. As discussed earlier, we would be using the dilaton Weyl multiplet for the construction of higher derivative terms in $N=2$ Poincar{\'e} supergravity. However it is not sufficient to give us the Einstein-Hilbert term. We would need the tensor multiplet for that. The tensor multiplet would also be needed as a compensating multiplet to break the $SU(2)$-R symmetry as we will discuss in Appendix-\ref{SU2breaking}. In that case, the tensor multiplet becomes a part of the gravity multiplet.
	
	Imposing the gauge fixing condition obtained in section-\ref{4} on the superconformal tensor multiplet action (\ref{tma}), we obtain the following bosonic action in Poincar\'e supergravity that contains the Einstein-Hilbert term in the Jordan frame.
	\begin{eqnarray}\label{tma1}
	e^{-1} \mathcal{L}_{TR} &=&\frac{1}{2}LR + L^{-1}  L^{ij}\cD_{a} \cD^{a} L_{ij} +  \frac{1}{8} L \Big( {\cF} \cdot {\cF}  +  {\cG} \cdot {\cG} \Big) + |G|^2 L^{-1}-E^2 L^{-1}\nn \\ 
	&& + \frac{L}{24} {\cH}^{abc}{\cH}_{abc} +2i  \tilde{E}^{ab} \cD_{a}\left(E_{b}L^{-1}\right) +\frac{i}{2}\varepsilon^{i j} L_{ki} \tilde{E}^{ab} L^{-1} (\hat{V}_{ab})\;^{k}\;_{j}  \nonumber\\
&&  -  L^{ij} \cD_{a}L_{ik} \cD^{a}L_{jl} L^{kl} L^{-3} - i \tilde{E}^{ab} \cD_{a} L_{ik} \cD _{b} L_{jl}  L^{kl} \varepsilon ^{ij} L^{- 3} + \text{h.c}\;. 
	\end{eqnarray}
	In order to arrive at the above action, we have used the expression for the composite gauge field $D$ (\ref{T}) in the super-Poincar{\'e} background which is:
	\begin{align}\label{D}
	D=\frac{1}{6}R+\frac{1}{24}\cH_{abc}\cH^{abc}+\frac{1}{8}\left(\cF\cdot\cF+\cG\cdot\cG\right)
	\end{align}
	In order to get the above composite expression for $D$, we would need the expression for the $U(1)$ R-symmetry gauge field in the super-Poincar{\'e} backrground which is:
	\begin{align}\label{U1}
	A_{a}&=-\frac{i}{12}\varepsilon_{abcd}\cH^{bcd}
	\end{align}
	In the subsequent sections, we would need the action of the Yang-Mills multiplet in the super-Poincar{\'e} background which we obtain from (\ref{yma}) by imposing the gauge fixing conditions as shown below:
	\begin{eqnarray}\label{yma1}
	\Big(e^{-1} \mathcal{L}_{YM}\Big)^{bosonic} &=&  8 {\cD}_a A^\a {\cD}^a \bar{A}^\b a_{\a \b } + \frac{2}{3} \varepsilon_{abcd} \cH^{bcd}  A^\a \overleftrightarrow{\cD}^a \bar{A}^\b  a_{\a \b}  \nn \\ &+&    A^\a \bar{A}^\b \Big( \cF \cdot \cF + \cG \cdot \cG \Big)  a_{\a \b} -  Y_{ij}^\a Y^{\b ij } a_{\a \b}  -  8  [ A, \bar{A}]^2 \nonumber\\ &+& 2\hat{\cB}^{- \a} \cdot \hat{\cB}^{-\b }a_{\a \b} -  A^\a \hat{\cB}^{\b} \cdot T^{+} a_{\a \b}  + \text{h.c.}  
	\end{eqnarray}



	
	\subsection{Riemann Squared Action in pure $N=2$ supergravity}\label{5}
	The construction of Riemann squared invariant for minimal supergravity theories in five dimensions and six dimensions relied on the mapping between the Yang-Mills multiplet and a multiplet constructed out of the dilaton Weyl multiplet that contains a spin connection with bosonic torsion and a set of curvatures \cite{Ozkan:2013uk,Ozkan:2013nwa,BERGSHOEFF198673}. We would expect the similar mapping to exist in four dimensions which we will derive in the following subsection.

	\subsection{Map between Yang-Mills and Poincar\'e Multiplet} \label{map}
	
	Using the results for the SuperPoincar{\'e} transformation rules of the curvatures and spin connection in the dilaton Weyl multiplet obtained in section-\ref{4}, we get the following:
	\bea \label{ct}
	\delta{{K}}_{ab} &=& 4 \varepsilon_{ij} \bar{\epsilon}^{i} \widehat{\psi}_{ab}^j 
	,\nn \\
	\delta \widehat{\psi}_{ab}{}^i &=& - \frac{1}{2} \gamma^{cd} \widehat{R}_{cdab}(\omega_{-})\epsilon^{i} + (\widehat{V}_{ab})^i\;_j \epsilon^j -\frac{\varepsilon^{ij}}{2} \slashed{{D}}(\omega_{-}) {{K}}_{ab} \epsilon_{j} - \frac{1}{4}{\mathcal{F}}_{ab} \gamma \cdot {\mathcal{F}} \epsilon^{i}  \nn\\ && - \frac{1}{4}{\mathcal{G}}_{ab} \gamma \cdot {\mathcal{G}} \epsilon^{i} + \frac{i}{4} \eta^{cd} [{\mathcal{G}}_{ac}{\mathcal{F}}_{db} + {\mathcal{F}}_{ac}{\mathcal{G}}_{db}],\nn \\
	\delta {\omega}_{- \m ab} &=& \bar{\epsilon}_{i}\gamma_{\m}\widehat{\psi}_{ab}^{i} - \frac{1}{2} \varepsilon^{ij}\bar{\epsilon}_{i} {{ K}}_{ab} \psi_{\m j}, \nn\\
	\delta (\widehat{V}_{ab})_{ij} &=& 2\bar{\epsilon}_{(i} \varepsilon_{j)k} \slashed{D}(\o ,\o_-) \widehat{\psi}_{ab}{}^k + i \bar{\epsilon}_{(i}\varepsilon_{j) k} \gamma^c A_{c} \widehat{\psi}_{ab}{}^k - 2 \bar{\epsilon}_{(i}\;\widehat{\psi}^c\;_{[a\;j)} {K}_{b\;] c} + \text{h.c.} 
	\eea
	
	\noindent
	The supersymmetry transformations of Yang-Mills multiplet in Poincar\'e supergravity are given as follows.
	\begin{eqnarray}\label{ygmt}
	\delta A^{\a} &=& \bar{\epsilon}^{i} \lambda_{i}{}^{\a}   \ ,
	\nonumber \\
	\delta \lambda_{i}^{\a} &=& 2 \slashed{D} A^{\a} \epsilon_{i} +\frac{1}{2} \gamma \cdot \mathcal{B}^{ \a-} \varepsilon_{ij} \epsilon^{j} + Y_{ij}^{\a} \epsilon^{j} - \frac{\varepsilon_{ij}}{4} A^{\a} \gamma \cdot (\mathcal{F}^- - i \mathcal{G}^-) \epsilon^{j} \nn \\ 
	&& - \frac{\varepsilon_{ij}}{4} \bar{A}^{\a} \gamma \cdot (\mathcal{F}^- + i \mathcal{G}^-) \epsilon^{j} + A^{\b} \bar{A}^{\g} f_{\b \g}{}^{\a}  \varepsilon_{ij} \epsilon^{j} \ ,
	\nonumber\\
	\delta C_{\mu}^{\a} &=&  \varepsilon^{ij} \bar{\epsilon}_{i} \gamma_{\mu} \lambda_{j}{}^{\a} + 2\;\varepsilon^{ij} \bar{\epsilon}_{i} \psi_{\mu j} A^{\a} + \text{h.c}  \ ,
	\nonumber\\
	\delta Y_{ij}^\a &=& 2 \bar{\epsilon}_{(i} \slashed{D} \lambda_{j)}{}^\a 	 + i \bar{\epsilon}_{(i} \gamma^\m A_\m \lambda_{j)}^\a  + 2 \varepsilon_{m (i} \bar{\epsilon}_{j)} A^{\b} \lambda^{m \g} f_{\b \g}{}^\a - \text{h.c.}
	\end{eqnarray}
	where $D_{\mu}$ denotes the fully covariant derivative with respect to all Super-Poincar{\'e} transformations as mentioned below:
	\begin{eqnarray*}
		D_{\mu}A^\a &=& \partial_{\mu} A^{\a} + C_{\mu}^{\b} A^{\g} f_{\b \g}{}^{\a} - \frac{1}{2}\bar{\psi}_{\mu}{}^{i}\lambda_{i}^\a,
		\nonumber\\
		D_{\mu} \lambda_{j}{}^\a 	&=& \left( \partial_{\m} - \frac{1}{4} \omega_{\m}\;^{ab} \gamma_{ab} \right) \lambda_j^\a - \frac{1}{2} \mathcal{V}_{\m}^k\;_{j} \lambda_{k}{}^\a +  C_{\m}^\b \lambda_{j}^\g f_{\b \g}{}^\a - \slashed{D} A^\a \psi_{\m j} - \frac{1}{4} \varepsilon_{jk} \gamma \cdot \mathcal{B}^\a \psi_{\m}{}^k  \nonumber\\ && - \frac{1}{2} Y_{jk}^\a \psi_{\m}{}^k + \frac{1}{8} A^\a \varepsilon_{jk} \gamma \cdot (\mathcal{F}^- - i\mathcal{G}^-) \psi_{\m}{}^k + \frac{1}{8} \bar{A}^\a \varepsilon_{jk} \gamma \cdot (\mathcal{F}^- + i\mathcal{G}^-) \psi_{\m}{}^k\nn \\
		&&-\frac{1}{2}A^{\b}\bar{A}^{\g}f_{\b\g}{}^{\a}\varepsilon_{jk}\psi_{\mu}^{k}.
	\end{eqnarray*}
	We observe that the transformation rules in equations (\ref{ygmt}) and (\ref{ct}) become identical
	upon identifying the gauge group in the Yang-Mills multiplet with the Lorentz group and making the following identifications of the components:
	
	\begin{center}
		$\begin{pmatrix}
		A^\a \vspace{.3cm}\\
		
		\lambda_i{}^\a\vspace{.2cm}\\
		
		C_{\m}^\a\vspace{.3cm}\\
		
		Y_{ij}{}^\a
		\vspace{.3cm}
	\end{pmatrix} \longleftrightarrow \begin{pmatrix}
		\frac{{{K}}^{ab}}{4}
		\vspace{.3cm}\\
		
		\varepsilon_{ij} \widehat{\psi}^{ab j}\vspace{.2cm}\\
		
		- \omega_{-\m}{}^{ab}\vspace{.3cm}\\
		
		\varepsilon_{ik} (\widehat{V}{}^{ab})^k{}_{j}
		\vspace{.3cm}
	\end{pmatrix}$
	\end{center}	
	\noindent 
	Using this observation we can now easily write down a supersymmetric Riemann tensor squared action by using the Poincar\'e supergravity action for the Yang-Mills multiplet given in  Eq (\ref{yma1}). The bosonic part of the action is given as:
	\begin{eqnarray}\label{rsa}
	\Big(e^{-1} \mathcal{L}_{YM}\Big)^{bosonic}  &=& 4{\cD}_\m(\o_-) K^{ab} {\cD}^{\m}(\o_-) \bar{K} _{ab}  + \frac{1}{3}  \varepsilon_{abcd} \cH^{bcd} K_{ef} \overleftrightarrow{\cD}^a(\o_-) \bar{K}^{ef}  \nn \\ 
	&+&  16 R_{abcd}(\o_-)R^{abcd}(\o_-) -  (V_{ab})^{i}{}_{j}(V^{ab})^{j}{}_{i} 
	\nn \\
	&+& \frac{1}{2} K^{ab} \bar{K}_{ab} \Big( \cF \cdot \cF + \cG \cdot \cG \Big)  - 4 K^{ab} R_{cdab}{}(\o_-) \bar{K}^{cd+}  + \text{h.c.} 
	\end{eqnarray}
	\noindent
	Since the Yang-Mills group has been mapped to Lorentz group, while computing the above action we have replaced the trace of Yang-Mills group $a_{\a \b} = Tr(T_\a T_\b)$  with the trace $Tr(M_{[ab]}M_{[cd]}) = 4 \Big( \eta_{ac} \eta_{bd} - \eta_{ad}\eta_{bc} \Big)$ of the Lorentz group.
	\section{Vector multiplets coupled higher derivative supergravity}\label{6}
	In the previous section, we saw how we can obtain supersymmetrization of Riemann square term in pure $N=2$ Poincar{\'e} supergravity. However to make our results applicable to study of black holes it would be desirable to couple arbitrary number of vector multiplets and obtain supersymmetrization of arbitrary curvature squared terms. In order to obtain supersymmetrization of arbitrary curvature square terms, we make the following observations
	\begin{enumerate}
	\item
	From our discussions in the previous section, one can obtain the supersymmetrization of Riemann square term by starting from the Yang-Mills action and mapping some combination of the fields belonging to the dilaton Weyl multiplet to the Yang-Mills multiplet with Lorentz group as the gauge group. The Yang-Mills action on the other hand is obtained from the chiral density formula by mapping the lowest component of the chiral multiplet to the gauge invariant combination $A^{\alpha}A^{\beta}Tr(T_{\alpha} T_{\beta})$.
	\item
	Supersymmetrization of a particular combination of the Weyl square and Ricci scalar square term can be obtained by embedding the dilaton Weyl multiplet inside the chiral multiplet via the identification of $ T^{-}\cdot T^{-}$ with the lowest component of the chiral multiplet and using it in the chiral density formula. Similar construction exists for the standard Weyl multiplet \cite{Bergshoeff:1980is} where one obtains the supersymmetrization of Weyl square term. However, since in dilaton Weyl multiplet the field $D$ is a composite field unlike the similar field $D$ in the standard Weyl multiplet, there will be an additional Ricci scalar square term coming from the term proportional to $D^2$ term in the action.
	\item
	Supersymmetrization of Ricci scalar square term can be obtained by embedding the tensor multiplet inside the chiral multiplet by identifying $B^2$ with the lowest component of the chiral multiplet, where $B$ is the composite defined in (\ref{m}) that transforms as the lowest component of a vector multiplet. This observation arises from the fact that the action constructed in this way will have a term $\mathcal{Z}_{ij}\mathcal{Z}^{ij}$ in the action where $\mathcal{Z}_{ij}$ is the composite defined in (\ref{m}) and contains a factor of Ricci scalar inside the superconformal de-Alembertian of $L_{ij}$.
	\end{enumerate}
	 So to achieve our desired results, we follow the following step-wise procedure:
	\begin{itemize}
	\item
	Step-1: We construct a chiral multiplet in conformal supergravity whose lowest component is given as:
	\begin{align}
	\hat{A}=F(X^I,X,B,\mathcal{A},\cP)
	\end{align}
         where $F$ is the prepotential and it depends holomorphically on the arguments inside the bracket which are:.
         \begin{enumerate}
         \item
          $X^I$ ($1\leq I\leq n_v$): the complex scalars of the $n_v$ vector multiplets which we want to couple to our theory.
          \item
          $X$: the complex scalar of the dilaton Weyl multiplet.
          \item
          $B$: the composite constructed out of tensor multiplet defined in (\ref{m}).
          \item
           $\cA=Tr(A^2)=A^{\alpha}A^{\b}a_{\a\b}$ where $A^{\a}$ are the complex scalars of the Yang-Mills multiplet and $a_{\alpha\beta}=Tr(T_{\alpha}T_{\beta})$, $T_{\alpha}$ being the generators of the Yang-Mills gauge group.
           \item
           $\cP=T^{-}\cdot T^{-}$, where $T^{-}_{ab}$ is given as a composite in terms of the dilaton Weyl multiplet in (\ref{T}) and transforms like the $T^{-}_{ab}$ field of the standard Weyl multiplet.
           \end{enumerate}
           The prepotential also needs to be a homogeneous function degree of 2 i.e.,
	\begin{align}
	F(\Lambda X^I,\Lambda X, \Lambda B, \Lambda^2 \cA,\Lambda^2 \cP) = \Lambda^2 F(X^I, X, B, \cA,\cP)
	\end{align}
	\item
	Step-2: Use the chiral density formula to construct an invariant action out of the above chiral multiplet.
	\item
	Step-3: Apply Poincar{\'e} gauge fixing conditions discussed in section-\ref{4} and Yang-Mills map discussed in section-\ref{map} to obtain the desired action.
	\end{itemize}
	\noindent
	The other components of the chiral multiplet are obtained by the application of subsequent supersymmetry transformations. We obtain:
	\begin{eqnarray}\label{ccf}
	\hat{\Psi}_i &=& F_I \Omega_{i}^I + F_{X} \Omega_{i} + 2 F_{\mathcal{A}}A^\a \lambda_i^\b a_{\a \b} + F_\cP \big( 16\varepsilon_{ij} R(Q)_{ab}^j T^{-ab}\big)+ F_B \Sigma_i\;, 
 \nn \\ 
	\hat{B}_{ij} &=& - \frac{1}{2} F_{IJ} \bar{\Omega}_{i}^I \Omega_{j}^J - \frac{1}{2}F_{XX} \bar{\Omega}_{i} \Omega_{j} - 2 F_{\mathcal{A} \mathcal{A}} A^\g A^\a \bar{\lambda}_{i}^\b \lambda_{j}^\d a_{\a \b}a_{\g \d} - F_{\cA}\bar{\lambda}_{i}^\a \lambda_{j}^\b a_{\a \b} \nn\\
	&&  - F_{XI} \bar{\Omega}_{(i}^I \Omega_{j)} - 2 F_{\cA I}A^\a \bar{\Omega}_{(i}^I \lambda_{j)}^\b a_{\a \b}  - 128 \varepsilon_{ik}\varepsilon_{jl}F_{\cP \cP}T^{-ab}T^{-cd}\bar{R}_{ab}(Q)^k{R}_{cd}(Q)^l 
	\nn \\
	&&-  2 F_{\cA X}A^\a \bar{\Omega}_{(i} \lambda_{j)}^\b a_{\a \b} + F_{I} Y_{ij}^I + 2 F_{\cA} A^\a Y_{ij}^\b a_{\a \b}  - 64 \varepsilon_{ik}\varepsilon_{jl}F_{\cP }\bar{R}_{ab}(Q)^k{R}^{ab}(Q)^l 
	\nn \\ 
	&&- 16 F_{\cP} T^{-ab} \varepsilon_{k(i}R(V)^k\;_{j)ab} +16 F_{X\cP} T^{-ab} \varepsilon_{k(i}\bar{\Omega}_{j)}{R}_{ab}(Q)^k 
	\nn \\  
	&&+  16 F_{I\cP} T^{-ab} \varepsilon_{k(i}\bar{\Omega}^I_{j)}{R}_{ab}(Q)^k + 32 F_{\cA\cP} A^{\a}T^{-ab} \varepsilon_{k(i}\bar{\lambda}^\b_{j)}{R}_{ab}(Q)^k a_{\a\b}
	\nn \\
	&&  + F_B \mathcal{Z}_{ij}  - \frac{1}{2}	F_{BB} \bar{\Sigma}_i \Sigma_j - F_{BI} \bar{\Sigma}_{(i} \Omega_{j)}^I - F_{BX} \bar{\Sigma}_{(i} \Omega_{j)} - 2F_{B \cA} \cA^\a \bar{\Sigma}_{(i} \lambda^\b_{j)} a_{\a\b} 
\nn \\
&&+16 F_{B\cP} T^{-ab} \varepsilon_{k(i}\bar{\Sigma}_{j)}{R}_{ab}(Q)^k \nn\\
	\widehat{F}_{ab}^- &=& F_{I} \hat{\cF}_{ab}{}^{-I} + \frac{1}{2} F_{X} ( \cF^- - i\cG^-) + 2 F _{\cA} A^{\a} \hat{\cB}_{ab}{}^{-\b}a_{\a\b} - \frac{1}{4} F_{\cA} \varepsilon^{kl} \bar{\lambda}_k^\a \gamma_{ab} \lambda_{l}^\b a_{\a \b} \nn \\ 
	&&- \frac{1}{8} F_{IJ} \varepsilon^{kl} \bar{\Omega}_k^I \gamma_{ab} \Omega_{l}^J - \frac{1}{8} F_{XX} \varepsilon^{kl} \bar{\Omega}_k \gamma_{ab} \Omega_{l} - \frac{1}{2} F_{\cA \cA} \varepsilon^{kl}A^\a A^\g \bar{\lambda}_k^\b \gamma_{ab} \lambda_{l}^\d a_{\a \b}a_{ \g \d}
	\nn \\
	&&- 32\varepsilon_{ij}F_{\cP \cP}T^{-ef}T^{-cd}\bar{R}_{ef}(Q)^k \gamma_{ab}{R}_{cd}(Q)^l  - 16 \varepsilon_{kl} F_{\cP} \bar{R}_{cd}(Q)^k \gamma_{ab}{R}_{cd}(Q)^l   \nn \\ 
	&&- 16 F_{\cP} T^{-cd} \cR_{cdab}(M) - 4F_{\cP I} T^{-cd} \bar{\Omega}_i^I \gamma_{ab} R_{cd}(Q)^i - 4F_{\cP X} T^{-cd} \bar{\Omega}_i \gamma_{ab} R_{cd}(Q)^i  
	\nn \\ 
	&&- 8 F_{\cP \cA} A^{\a} T^{-cd} \bar{\lambda}^\b_i \gamma_{ab}R_{cd}(Q)^i a_{\a\b}
	- \frac{1}{2} F_{I\cA} A^\a \varepsilon^{kl} \bar{\Omega}_k^I \gamma_{ab} \lambda_{l}^\b a_{\a \b} - \frac{1}{4} F_{XI} \varepsilon^{kl} \bar{\Omega}_k^I \gamma_{ab} \Omega_{l} \nn \\
	&&- \frac{1}{2} F_{X\cA} A^\a \varepsilon^{kl} \bar{\Omega}_k \gamma_{ab} \lambda_{l}^\b a_{\a\b}  + F_C \hat{\cC}_{ab}^-  - \frac{1}{8} F_{BB} \varepsilon^{kl} \bar{\Sigma}_k \gamma_{ab} \Sigma_{l} - \frac{1}{4} F_{BI} \varepsilon^{kl} \bar{\Omega}_k^I \gamma_{ab} \Sigma_{l} \nn \\
	&&- \frac{1}{4} F_{BX} \varepsilon^{kl} \bar{\Omega}_k \gamma_{ab} \Sigma_{l}
  - \frac{1}{2} F_{B\cA} A^\a \varepsilon^{kl} \bar{\Sigma}_k \gamma_{ab} \lambda_{l}^\b a_{\a\b} - 4F_{\cP B} T^{-cd} \bar{\Sigma}_i \gamma_{ab}R_{cd}(Q)^i  ,
	\nn \\
	2 \hat{\Lambda}_i &=& -\frac{1}{2} F_{IJ} \gamma \cdot \hat{\cF}^{-I} \Omega_{i}^J - \frac{1}{4} F_{XX}\gamma \cdot (\cF^- - i \cG) \Omega_i -  2 F_{\cA \cA} A^\g A^\a \gamma \cdot \hat{\cB}^{-\d} \lambda_{i}^\b a_{\a\b}a_{\g\d} 
	\nn \\ 
	&&- F_{\cA I} A^\a \gamma \cdot \hat{\cF}^{-I}\lambda_i^\b a_{\a \b} -  \frac{ F_{\cA X}}{2} A^{\a} \gamma \cdot ( \cF^- - i\cG^-)\lambda_{i}^\b a_{\a \b} - \frac{F_{IX}}{2} \gamma \cdot \hat{\cF}^{-I} \Omega_{i} 
	\nn \\
	&&- \frac{F_{IX}}{4}  \gamma \cdot ( \cF^- - i\cG^-)\Omega_i^I - F_{\cA X} A^\a \big(\gamma \cdot \hat{\cB}^{-\b} \big) \Omega_i a_{\a\b} - F_{\cA I} A^\a \big(\gamma \cdot \hat{\cB}^{-\b} \big) \Omega_i^Ia_{\a\b}
	\nn \\
	&&  - F_{\cA} \big( \gamma \cdot \hat{\cB}^{-\a}\big) \lambda_{i}^\b a_{\a \b} + 256  F_{\cP \cP}   T^{-ab} T^{-cd} \varepsilon_{ij} R_{ab}(V)^j{}_k R_{cd}(Q)^k
	\nn \\
	&& + 128 F_{\cP} \varepsilon  _{ij} R_{ab}(V)^j{}_k R^{ab}(Q)^k  + 128 F_{\cP \cP}  T^{-ab}T^{-cd} \gamma_{ef} \varepsilon_{ij} R_{ab}(Q)^j \tilde{\cR} (M)_{cd}{}^{ef}
	\nn \\
	&&+  64 F_{\cP} \varepsilon_{ij} \gamma_{ab} R_{cd}(Q)^j \tilde{\cR}(M)^{cd ab} + 32 F_\cP T^{-ab} \Big( R_{ab}(S)_i + 3 \gamma_{[a}\cD_{b]} \chi_i \Big) 
	\nn \\
	&&+ 8 F_{I\cP} \gamma_{ab} T^{-cd} {\cR}(M)^{ab}{}_{cd} \Omega_i^I + 8 F_{X \cP}  \gamma_{ab} T^{-cd} {\cR}(M)^{ab}{}_{cd} \Omega_i 
	\nn\\ 
	&&+  16 F_{\cP \cA} A^{\a} \gamma_{ab} T^{-cd} {\cR}(M)^{ab}{}_{cd} \lambda_i^\b a_{\a\b}  - 8 F_{I\cP} \varepsilon_{ij} T^{-ab} \gamma \cdot \hat{\cF}^{-I} R_{ab}(Q)^j  
	\nn\\ 
	&& - 8 F_{X \cP} \varepsilon_{ij} T^{-ab} \gamma \cdot \Big( {\cF}^- -i \cG^-\Big)  R_{ab}(Q)^j  - 16 F_{\cA \cP} A^{\a} \varepsilon_{ij} T^{-ab} \gamma \cdot \hat{\cB}^{-\b} R_{ab}(Q)^j a_{\a\b}
	\nn \\
	&& + 16 F_{I\cP} T^{-ab} Y_{ij}^I R_{ab}(Q)^j + 32 F_{\cA\cP} A^\a T^{-ab} Y_{ij}^\b \cR_{ab}(Q)^j a_{\a\b} + F_{IJ} \varepsilon^{lm} \Omega_{l}^I Y_{mi}^J 
	\nn \\ 
	&&+ F_{IX} \varepsilon^{lm} \Omega_l Y_{mi}^I + 4  F_{\cA \cA} \varepsilon^{lm} A^\g A^\a \lambda_l^\b Y_{mi}^\d a_{\a\b}a_{\g \d} + 2 F_\cA \varepsilon^{lm} \lambda_l^\a Y_{mi}^\b a_{\a \b}
	\nn \\
	&&-  16 F_{I\cP} T^{-ab} R_{ab}(V)^j{}_i \Omega_j^I - 16 F_{X\cP} T^{-ab} R_{ab}(V)^j{}_i \Omega_j 
	\nn \\
	&& - 32 F_{\cA \cP} T^{-ab} \varepsilon_{ij} \varepsilon^{kl} A^\a \cR_{ab}(V)^j{}_k \lambda_l^\b a_{\a\b}  - 4 F_\cA f_{\g \d}{}^\a A^\g \bar{A}^\d \lambda_{i}^\b a_{\a \b}
	\nn \\
	&&   + 2 F_{I\cA} \varepsilon^{lm} A^\a \lambda_l^\b Y_{mi}^I a_{\a \b} +   2 F_{\cA X} \varepsilon^{lm} A^\a \Omega_l Y_{mi}^\b a_{\a \b}  + 2 F_{I\cA}\varepsilon^{lm} A^\a \Omega_l^I Y_{mi}^\b a_{\a \b} 
	\nn \\
	&& + 2 F_{I} \Gamma^{m I} \varepsilon_{mi} + 4 F_{\cA} A^\a \Pi ^{m\b} \varepsilon_{mi} a_{\a\b} + 2 F_B \Gamma^{m} \varepsilon_{mi} + F_{BB} \varepsilon^{lm} \Sigma_l \cZ_{mi}	 \nn\\
	&&  + F_{IB} \varepsilon^{lm} \Sigma_l Y_{mi}^I + F_{IB} \varepsilon^{lm} \Omega_l ^I \cZ_{mi}  + F_{XB} \varepsilon^{lm} \Omega_l \cZ_{mi} + 2 F_{B\cA}\varepsilon^{lm} A^\a \Sigma_l Y_{mi}^\b a_{\a \b}  
\nn \\
&& + 2 F_{B\cA}\varepsilon^{lm} A^\a \lambda_l ^\b \cZ_{mi} a_{\a \b}  +  16 F_{B\cP} T^{-ab} \ R_{ab}(Q)^j \cZ_{ij}  -\frac{1}{2} F_{BB} \gamma \cdot \hat{\cC}^{-} \Sigma_{i} 
	\nn \\
	&&  - \frac{1}{2} F_{BI} \gamma \cdot \hat{\cF}^{-I} \Sigma_i - \frac{1}{2} F_{BI} \gamma \cdot \hat{\cC}^- \Omega_i^I - \frac{1}{4} F_{BX} \gamma \cdot \Big( \cF^- - i \cG^- \Big) \Sigma_i - \frac{1}{2} F_{BX} \gamma \cdot \hat{\cC}^- \Omega_i 
	\nn \\
	&&- F_{B \cA} A^\a \gamma \cdot \hat{\cB}^{-\b} \Sigma_i a_{\a \b} - F_{B\cA} A^\a \gamma \cdot \hat{\cC}^- \lambda_i^\b a_{\a \b} - 8 F_{B\cP} \varepsilon_{ij}T^{- ab} \gamma \cdot \hat{\cC}^- R_{ab}(Q)^j 
\nn \\	
	&&  + 8 F_{B \cP} \gamma_{ab} T^{-cd} {\cR}^{ab}\;_{cd}(M) \Sigma_i - 16 F_{B \cP}  T^{-ab} R_{ab}(V)^j\;_i \Sigma_j 
	\nn  \\
	2 \hat{C} &=& - 4 F_{I} {\square}^c \bar{X}^I + 4 F_{I} \bar{X}^I D  - \frac{1}{2} F_{I} \hat{\cF}^I \cdot T^+ - 4 F_{B} {\square}^c \bar{B} + 4 F_{B} \bar{B} D  - \frac{1}{2} F_{B} \hat{\cC} \cdot T^+ 
	\nn \\&&  - 8 F_{\cA} A^\a \square^c  \bar{A}^\b a_{\a\b}+8F_{\cA} A^\a \bar{A}^\b Da_{\a\b}  - F_{\cA} A^\a \hat{\cB}^\b \cdot T^+ a_{\a\b} + F_{IJ} \hat{\cF}^-_{ab}{}^{I} \hat{\cF}^{-abJ} \nn \\
	&&+ \frac{1}{4} F_{XX} (\cF^{-ab}-i\cG^{-ab})^2 + F_{BB} \hat{\cC}^{-ab} \hat{\cC}^-_{ab}  - \frac{1}{2}F_{BB} \cZ_{ij} \cZ^{ij} - F_{BI} Y_{ij}^I \cZ^{ij} \nn \\
	&& + 2  F_{\cA} \hat{\cB}^{-\a ab}\hat{\cB}^{-\b}_{ab}a_{\a\b} + 2 F_{BI} \hat{\cC}^{-ab} \hat{\cF}_{ab}^{-I} + F_{BX} \Big(\cF^{-ab} -i \cG^{-ab}\Big)  \hat{\cC}_{ab}^-
	\nn \\
	&&+ 4 F_{\cA \cA} A^\g A^\a \hat{\cB}^{-\b ab}\hat{\cB}^{-\d}_{ab}a_{\a\b}a_{\g\d} + 4 F_{I\cA} A^\a \hat{\cF}^{-Iab}\hat{\cB}^{-\b}_{ab}a_{\a\b} -  F_{\cA} Y_{ij}^\a Y^{\b ij} a_{\a \b}
	\nn \\
	&& +2F_{\cA X} A^\a (\cF^{-ab} - i \cG^{-ab})\hat{\cB}^{-\b}_{ab} a_{\a\b} + F_{I X}  (\cF^{-ab} - i \cG^{-ab})\hat{\cF}^{-I}_{ab} - \frac{1}{2} F_{IJ} Y_{ij}^I Y^{J ij} 
	\nn \\  
	&& - 2F_{\cA I} A^\a Y_{ij}^I Y^{\b ij} a_{\a\b} - 8 F_{\cA} [ A , \bar{A}]^2 +  128 F_{\cP}  {\cR}(M)_{cd}{}^{ ab} {\cR}(M)^{- cd }{}_{ab} 
	\nn \\
	&&+ 256 F_{\cP \cP}  T^{-ab}T^{-cd}  {\cR}(M)^-_{ab ef} {\cR}(M)_{cd}{}^{ ef} +   64 F_{\cP}  {R}(V)_{ab}^{-k}{}_{l} {R}(V)^{ab l}{}_k 
	\nn \\ 
	&&+ 128 F_{\cP \cP}  T^{-ab}T^{-cd}  {R}(V)_{ab}^{-k}{}_{l} {R}(V)_{cd}^{-l}{}_{k}  - 32F_\cP T^{-ab} D_a D^c T^{+}_{cb} 
	\nn \\
	&&- 32 F_{X \cP}  T^{-ab} {\cR}(M)_{cdab} \Big( {\cF}^{-cd} -i \cG^{-cd}\Big)  -  32 F_{I \cP}  T^{-ab} {\cR}^{cd}{}_{ab}(M) \hat{\cF}^{-I}_{cd} 
	\nn \\
	&& - 64 F_{\cA \cP}  T^{-ab} A^\a {\cR}^{cd}{}_{ab}(M) \hat{\cB}^{-\b}_{cd} a_{\a\b} - 16 F_{I \cP} T^{-ab} \varepsilon^{ij} R_{ab}(V)^k{}_i Y_{jk}^I 
	\nn \\  
	&&  -32  F_{\cA \cP} T^{-ab} A^\a\varepsilon^{ij} R_{ab}(V)^k{}_i Y_{jk}^\b a_{\a\b} - 16 F_{B \cP} T^{-ab} \varepsilon^{ij} R_{ab}(V)^k{}_i \cZ_{jk}
	\nn \\
	&&   - 2F_{B\cA} A^\a Y_{ij}^\b \cZ^{ij} a_{\a \b} + 4 F_{B\cA} A^\a \hat{\cC}^{-ab} \hat{\cB}_{ab}^{-\b} a_{\a\b} -  32 F_{B \cP}  T^{-ab} {\cR}^{cd}{}_{ab}(M) \hat{\cC}^{-}_{cd} 
	\nn \\
	\end{eqnarray}
	Here, we have omitted the terms trilinear in fermions in the expression of $\hat{\Lambda}_i$ and all fermionic terms in  $\hat{C}$. The modified superconformal curvature ${\cR}(M)_{ab}{}^{cd} $ for local Lorentz transformation appearing above is defined as:
	\begin{align}\label{mfs}
	{\cR}(M)_{ab}{}^{cd}  = {R'}(M)_{ab}{}^{cd} + \frac{1}{32} \Big(T_{cd}^{-} T_{ab}^{+}  + T_{ab}^{-} T_{cd}^{+} \Big) 
	\end{align}
	where ${R'}(M)_{ab}{}^{cd}$ differs from the fully supercovariant curvature ${R}(M)_{ab}{}^{cd}$ (\ref{supcurv}) for the local Lorentz transformation via a field redefinition of the $K$-gauge field $f_{\mu}^{a}$. More precisely:
	\begin{align}
        {R'}(M)_{\m \n}{}^{ab} = R_{\m \n}{}^{ab} - 4 {f'}_{[\m}{}^{[a} e_{\n]}{}^{b]} + \text(fermionic\; terms)\;.
        \end{align}
        Where $R_{\m\n}{}^{ab}=2\partial_{[\m}\o_{\n]}{}^{ab}-2\o_{[\m}{}^{ac}\o_{\n]c}{}^{b}$ is the Riemann tensor at the bosonic level and ${f'}_{\mu}{}^{a}$ differs from $f_{\mu}^{a}$ (\ref{dependent}) as shown below\footnote{Note that the ${f'}_{\mu}{}^{a}$ is exactly the form of the K-gauge field that appears in the standard Weyl multiplet with the conventional set of constraints. The conventional set of constraints gets modified for the dilaton Weyl multiplet which changes the form of the K-gauge field $f_{\mu}{}^{a}$ as it appears in (\ref{dependent})}:
	\be \label{cf}
	{f'}_\m{}^a = \frac{1}{2} {R}_\m{}^a - \frac{1}{4}\big(D + \frac{1}{3}{R}\big) e_\m{}^a - \frac{i}{2} \tilde{R}_\m{}^a(A) + \frac{1}{32} T_{\m b}^{-}T^{ab+}
	\ee
	Substituting the expression of the modified K-gauge field (\ref{cf}) in (\ref{mfs}), we find that the bosonic part of ${\cR}_{\m\n}{}^{ab}$ is given as:
	\begin{eqnarray}
	{\cR}(M)_{\m\n}{}^{ab} &=& \Big( R_{\m \n}{}^{ab} - 2e_{[\m}{}^{[a} R_{\n]}{}^{ b]} + \frac{1}{3} R e_{[\m}{}^{[a} e_{\n]}{}^{ b]} \Big) + 2i \tilde{R}(A)_{[\m}{}^{[a}e_{\n]}{}^{b]}+ D e_{[\m}{}^{[a} e_{\n]}{}^{b]}
	\nn \\ 
	&=& C_{\m \n}{}^{ab}  + 2i \tilde{R}(A)_{[\m}{}^{[a}e_{\n]}{}^{b]} + D e_{[\m}{}^{[a} e_{\n]}{}^{b]}
	\end{eqnarray}
	\noindent 
	where $C_{\m\n}{}^{ab}$ is the Weyl tensor. Upon inserting the chiral multiplet obtained in (\ref{ccf}), we will obtain an action, the bosonic part of which can be obtained directly from the bosonic part of the highest component $\hat{C}$ of the chiral multiplet and we get:
	\begin{eqnarray}\label{swta}
	e^{-1}\cL &=&  4\cD'^{a} F_{I} \cD'_a \bar{X}^I +  4 A_{a}A^{a}F_{I}\bar{X}^{I}+4iA_{a}F_{I}\overleftrightarrow{\cD'}^{a}\bar{X}^{I} - 4 F_I \bar{X}^I f_{a}{}^a+ 4F_I \bar{X}^I D
	 \nn \\
	 && + 4\cD'^{a} F_{B} \cD'_a \bar{B} +  4 A_{a}A^{a}F_{B}\bar{B} + 4iA_{a}F_{B}\overleftrightarrow{\cD'}^{a}\bar{B} - 4 F_B \bar{B} f_{a}{}^a+ 4F_B \bar{B} \;D
	 \nn \\
	&&+  8 \cD'^{a}(F_{\cA} A^{\a}) \cD'_a \bar{A}^\b  a_{\a\b} +8A_{a}A^{a}F_{\cA}A^{\a}A^{\b}a_{\a\b}+8iA_{a}(F_{\cA}A^{\a})\overleftrightarrow{\cD'}^{a}A^{\b}a_{\a\b}
	\nn \\
	 && + 8 F_\cA A^\a \bar{A}^\b D a_{\a \b} -8 F_\cA A^\a \bar{A}^\b f_{a}{}^a a_{\a\b} - \frac{1}{2} F_{I} \hat{\cF}^I \cdot T^+  - \frac{1}{2} F_{B} \hat{\cC} \cdot T^+
	\nn \\
	&&- F_{\cA} A^\a \hat{\cB}^\b \cdot T^+ a_{\a\b}  + 2  F_{\cA} \hat{\cB}^{-\a ab}\hat{\cB}^{-\b}_{ab}a_{\a\b} + F_{IJ} \hat{\cF}^-_{ab}{}^{I} \hat{\cF}^{-abJ} + F_{BB} \hat{\cC}^{-ab} \hat{\cC}^-_{ab} 
	\nn\\
	&&-  F_{\cA} Y_{ij}^\a Y^{\b ij} a_{\a \b} - \frac{1}{2} F_{IJ} Y_{ij}^I Y^{J ij} - \frac{1}{2}F_{BB} \cZ_{ij} \cZ^{ij} - 2F_{\cA I} A^\a Y_{ij}^I Y^{\b ij} a_{\a\b}
	\nn \\
	&& - F_{BI} \cZ_{ij}^I \cZ^{ij}  - 2F_{B\cA} A^\a Y_{ij}^\b \cZ^{ij} a_{\a \b} + 4 F_{\cA \cA} A^\g A^\a \hat{\cB}^{-\b ab}\hat{\cB}^{-\d}_{ab}a_{\a\b}a_{\g\d} 
	\nn \\
	&& + 4 F_{I\cA} A^\a \hat{\cF}^{-Iab}\hat{\cB}^{-\b}_{ab}a_{\a\b} + 2 F_{BI} \hat{\cC}^{-ab} \hat{\cF}_{ab}^{-I} + F_{BX} \Big(\cF^{-ab} -i \cG^{-ab}\Big)  \hat{\cC}_{ab}^- 
	\nn \\
	&&  + 4 F_{B\cA} A^\a \hat{\cC}^{-ab} \hat{\cB}_{ab}^{-\b} a_{\a\b}  - 8 F_{\cA} [ A , \bar{A}]^2 +  128 F_{\cP}  {C}_{cd}{}^{ ab} {C}^{ - cd }{}_{ab}  + 384 F_\cP D^2 
	\nn \\
	&&  +  256 F_{\cP} {R}_{ab}(A) {R}^{- ab}(A) + 256 F_{\cP \cP}  T^{-ab}T^{-cd} {C}_{ab ef} {C}_{cd}{}^{ ef}  + 256 F_{\cP \cP}  \cP{D^2} 
	\nn \\
	&&+  512 F_{\cP \cP}  T^{-ab}T^{-cd} D C_{abcd} - 128  F_{\cP \cP}  \cP {R}(A) \cdot {R}(A) - 32 F_{B\cP}DT^{-ab}\hat{\cC}_{ab}
	\nn \\ 
	&&+512 F_{\cP\cP} T^{-ab} T^{-cd} {R}_{ac}(A) {R}(A)_{bd} + 1024  i F_{\cP\cP} T^{-ab}T^{-cd} C_{cdeb} {R}(A)_a{}^e 
	\nn \\
	&& - 64  F_{\cA \cP} D A^\a T^{-ab} \hat{\cB}^{-\b}_{ab} a_{\a\b} - 64 F_{\cA \cP}   T^{-ab} A^\a {C}^{cd}{}_{ab} \hat{\cB}^{-\b}_{cd} a_{\a\b} 
	\nn \\
	&&  + 128 i  F_{\cA \cP}  T^{-}_a{}^{b} A^\a \hat{\cB}^{-\b}_{bc} {R}(A)^{ca} a_{\a\b} + 128 F_{\cP \cP}  T^{-ab}T^{-cd}  {R}(V)_{ab}^{-k}{}_{l} {R}(V)_{cd}^{-l}{}_{k} 
	\nn \\
	&&+   64 F_{\cP}  {R}(V)_{ab}^{-k}{}_{l} {R}(V)^{ab l}{}_k + 64 i  F_{X \cP}  T^-_a{}^{b} \Big( {\cF}{}^{-}_{bc} -i \cG^{-}_{bc}\Big)  {R}(A)^{ca}  \nn\\
	&&+  32\cD'_{a}(F_\cP T^{- ab}) \cD'^c T^{+}_{cb}+32 F_{\cP}A_{a}A_{c}T^{-ab}T^{+c}{}_{b}+32iA_{a}(F_{\cP}T^{-ab})\overleftrightarrow{\cD'}_{c}T^{+c}{}_{b}
	\nn \\
	&&+32F_{\cP}f_{ac}T^{-ab}T^{+c}{}_{b}    -  32 F_{I \cP} T^{-ab} {C}^{cd}{}_{ab} \hat{\cF}^{-I}_{cd} + 64 i  F_{I \cP}  T^{-}_a{}^{b} \hat{\cF}^{-I}_{bc} \tilde{R}(A)^{ca} 
	\nn \\ 
	&& - 32  F_{I \cP} D T^{-ab} \hat{\cF}^{-I}_{ab}  -  16 F_{I \cP} T^{-ab} \varepsilon^{ij} R_{ab}(V)^k{}_i Y_{jk}^I -   32 F_{X \cP} D T^{-ab} \Big( {\cF}^{-}_{ab} -i \cG^{-}_{ab}\Big)  
	\nn \\ 
	&& -  32 F_{X \cP} T^{-ab} {C}_{cdab} \Big( {\cF}^{-cd} -i \cG^{-cd}\Big)   -32  F_{\cA \cP} T^{-ab} A^\a\varepsilon^{ij} R_{ab}(V)^k{}_i Y_{jk}^\b a_{\a\b}  
	\nn \\
	&&+ \frac{1}{4} F_{XX} (\cF^{-ab}-i\cG^{-ab})^2 + F_{I X}  (\cF^{-ab} - i \cG^{-ab})\hat{\cF}^{-I}_{ab} -  32 F_{B \cP}  T^{-ab} C^{cd}{}_{ab} \hat{\cC}^{-}_{cd}
	\nn \\
	&&   +64iF_{B\cP}T^{-}_{a}{}^{b}\hat{\cC}^{-}_{bc}{R}(A)^{ca} - 16 F_{A \cP} T^{-ab} \varepsilon^{ij} R_{ab}(V)^k{}_i Y_{jk}  \nn\\
	&& +2F_{\cA X} A^\a (\cF^{-ab} - i \cG^{-ab})\hat{\cB}^{-\b}_{ab} a_{\a\b}  + \text{h.c}.
	\end{eqnarray}
	In bringing the Lagrangian to the above form we have extracted the K-gauge field and the U(1) gauge field out of the covariant derivatives and performed some integration by parts. In the above equation we have denoted derivatives of the prepotential $F$ with respect to its arguments as defined in (\ref{prepotder}).
	
	\subsection{Poincar{\'e} gauge fixing}
	Imposing the gauge fixing condition obtained in subsection-\ref{4} on the superconformal action obtained above (\ref{swta}), we obtain an action involving a set of vector multiplets and a Yang-Mills multiplet in Poincar\' e Supergravity\footnote{We will consider the tensor multiplet to be a part of the Poincar{\'e} supergravity multiplet which is indeed the case once we break the $SU(2)$ R-Symmetry as discussed in Appendix-\ref{SU2breaking}}. We define the prepotential for Poincar{\'e} supergravity as a gauge fixed version of the prepotential defined for conformal supergravity i.e
	\begin{center}
		$F(X^I, X,B, \cA, \cP)_{|X=1} = G (X^I ,B, \cA, \cP) $
	\end{center}
	The only restriction on $G$ is that it is a holomorphic function of the arguments. The prepotential $F$ in conformal supergravity had a restriction on its Weyl and chiral weight and hence it had to be a homogenous function of degree $+2$ but once we break these symmetries by imposing $X=\bar{X}=1$, the prepotential $G$ in Poincare supergravity is devoid of any such restrictions. We would like to obtain the action purely in terms of the prepotential $G$ and its derivatives. We derive a set of identities between the derivatives of $G$ and the derivatives of $F$ after imposing the gauge fixing conditions in Appendix-\ref{A}. 
	\noindent 
	The bosonic part of the Lagrangian density in Poincar\' e supergravity theory is given below. For the sake of simplicity and making the Lagrangian easily readable, we decompose it into a part that involves supercovariant derivatives of the prepotential and we write the remaining part by decomposing it into various kinds of derivatives of the prepotential $G$ with respect to its arguments.
	
	\begin{eqnarray}\label{fl}
	e^{-1}\cL&=&4 \cD^a {G}_{I} \cD_a \bar{X}^I+\frac{1}{3}\varepsilon_{abcd}\cH^{bcd}G_{I}\overleftrightarrow{\cD}^{a}\bar{X}^{I}+ 8 \cD^{a}({G}_{\cA} A^{\a}) \cD_a \bar{A}^\b  a_{\a\b}+\nn\\
	&& \frac{2}{3}\varepsilon_{abcd}\cH^{bcd}(G_{\cA}A^{\a})\overleftrightarrow{\cD}^{a}A^{\b}a_{\a\b}+ 4\cD^a {G}_{B} \cD_a \bar{B}+\frac{1}{3}\varepsilon_{abcd}\cH^{bcd}G_{B}\overleftrightarrow{\cD}^{a}\bar{B}
\nn\\
	&& + 32\cD_{a}(G_\cP T^{- ab}) \cD^c T^{+}_{cb}-8\cH^{bgh}(G_{\cP}T^{-}_{gh})\overleftrightarrow{\cD}_{c}T^{+c}{}_{b} +G\cL_{0}+G_{I}\cL_{1}^{I}
	\nn \\
	&& +G_{\cA}\cL_{2} +G_{\cP}\cL_{3}+G_{\cA I}\cL_{4}^{I}+G_{\cP I}\cL_5^{I}+G_{IJ}\cL_6^{IJ}+G_{\cA\cA}\cL_{7}+G_{\cA\cP}\cL_{8}\nn\\
	&& +G_{\cP\cP}\cL_{9} + G_{B} \cL_{10} + G_{BB} \cL_{11} + G_{BI} \cL_{12}^{I} + G_{B \cA} \cL_{13}  +  G_{B \cP} \cL_{14} + \text{h.c.}
	\end{eqnarray}
	where,
	\begin{eqnarray}
	\cL_{0}&=&\frac{1}{2}\Big(\cF^{-ab}-i\cG^{-ab}\Big)^2
	\nn\\
	\cL_{1}^{I}&=&-\frac{1}{2}\hat{\cF}^{I}\cdot T^{+}+\frac{1}{2}\bar{X}^{I}\Big(\cF\cdot\cF+\cG\cdot\cG\Big)-\frac{1}{2}X^{I}\Big(\cF^{-ab}-i\cG^{-ab}\Big)^2+\Big(\cF^{-ab}-i\cG^{-ab}\Big)\hat{\cF}^{-I}_{ab}\nn\\
	\cL_{2}&=&A^{\a}\bar{A}^{\b}\Big(\cF\cdot\cF+\cG\cdot\cG\Big)a_{\a\b}-A^{\a}\hat{\cB}^{\b}\cdot T^{+}a_{\a\b}-\frac{1}{2}\cA \Big(\cF^{-ab}-i\cG^{-ab}\Big)^2 +2\hat{\cB}^{-\a}\cdot\hat{\cB}^{-\b}a_{\a\b}
	\nn\\
	&&   -Y^{\a}_{ij}Y^{\b ij}a_{\a\b}-8\left[A,\bar{A}\right]^2\nn\\
	\cL_{3}&=&-\frac{1}{2}\cP\Big(\cF^{-ab}-i\cG^{-ab}\Big)^2+128C_{abcd}C^{-abcd}+384D^{2}+64 \Big(V^{ab}\Big)^{i}{}_{j} \Big(V_{ab}\Big)^{-j}{}_{i}\nn\\
	&& +4\cH_{aef}\cH_{c}{}^{ef}T^{-ab}T^{+c}{}_{b} +16 R_{ab}T^{-ac}T^{+b}{}_{c}-32 \cD_{d}\cH^{abd}\cD^{e}\cH_{abe}\nn\\
	&&+16\varepsilon_{abcd} \cD_{f}\cH^{abf} \cD_{g}\cH^{cdg}\nn \\
	\cL_{4}^{I}&=& A^{\a}A^{\b}a_{\a\b}X^{I}\Big(\cF^{-ab}-i\cG^{-ab}\Big)^2+4A^{\a}\hat{\cF}^{-Iab}\hat{\cB}^{-\b}_{ab}a_{\a\b} -2A^{\a}Y^{I}_{ij}Y^{\b ij}a_{\a\b} \nn \\
	&& -2A^{\a}A^{\b}a_{\a\b}\Big(\cF^{-ab}-i\cG^{-ab}\Big)\hat{\cF}^{-I}_{ab} -2X^I A^{\a}\Big(\cF^{-ab}-i\cG^{-ab}\Big)\hat{\cB}^{-\b}_{ab}a_{\a\b}\nn\\
	\cL_{5}^{I}&=& T^{-}\cdot T^{-}X^{I}\Big(\cF^{-ab}-i\cG^{-ab}\Big)^2-2 T^{-}\cdot T^{-}\Big(\cF^{-ab}-i\cG^{-ab}\Big)\hat{\cF}^{-I}_{ab}\nn\\
	&& +32X^I DT^{-ab}\Big(\cF^{-}_{ab}-i\cG^{-}_{ab}\Big)-32T^{-ab}C_{abcd}\hat{\cF}^{-Icd}-32DT^{-ab}\hat{\cF}_{ab}^{-I} \nn\\
	&&+32 X^I T^{-ab}\Big(\cF^{-}_{cb}-i\cG^{-}_{cb}\Big)\cD_{d}\cH_{a}{}^{cd} +32X^I T^{-ab}C_{abcd}\Big(\cF^{-cd}-i\cG^{-cd}\Big)\nn \\
	&& -16T^{-ab}\varepsilon^{ij}\Big(V_{ab}\Big)^{k}{}_{i}Y_{jk}^{I} -32T^{-ab}\hat{\cF}^{-I}_{cb}\cD_{d}\cH_{a}{}^{cd} \nn \\
	\cL_{6}^{IJ}&=& \hat{\cF}^{-I}_{ab}\hat{\cF}^{-abJ}+\frac{1}{4}X^{I}X^{J}\Big(\cF^{-ab}-i\cG^{-ab}\Big)^{2}-\Big(\cF^{-ab}-i\cG^{-ab}\Big)X^{(I}\hat{\cF}^{-J)}_{ab}
	\nn \\
	&&-\frac{1}{2}Y^{I}_{ij}Y^{Jij}\nn\\
	\cL_{7}&=&4A^{\g}A^{\a}\hat{\cB}^{-\b}\cdot\hat{\cB}^{-\d}a_{\a\b}a_{\g\d}-4\cA{}\Big(\cF^{-ab}-i\cG^{-ab}\Big)A^{\g}\hat{\cB}^{-\d}_{ab}a_{\g\d}+\cA^2\Big(\cF^{-ab}-i\cG^{-ab}\Big)^{2} \nn\\ 
	\cL_{8}&=&2\cA{}\cP(\cF^{-ab}-i\cG^{-ab}\Big)^{2}-4\cP\Big(\cF^{-ab}-i\cG^{-ab}\Big)A^{\a}\hat{\cB}^{-\b}_{ab}a_{\a\b}-64DA^{\a}T^{-}\cdot\hat{\cB}^{-\b}a_{\a\b}\nn\\
	&&  -64T^{-ab}A^{\a}C_{abcd}\hat{\cB}^{-\b}{}^{cd}a_{\a\b}  -64T^{-ab}A^{\a}\hat{\cB}^{-\b}_{cb} \cD_{d}\cH_{a}{}^{cd}a_{\a\b} \nn\\
	&& +64 \cA T^{-ab}C_{abcd}\Big(\cF^{-cd}-i\cG^{-cd}\Big)+64\cA{}DT^{-ab}\Big(\cF^{-}_{ab}-i\cG^{-}_{ab}\Big)
	\nn \\
	&& -32\varepsilon^{ij} T^{ab-} \Big(V_{ab}\Big)^{k}{}_{i}A^{\a}Y^{\b}_{jk}a_{\a\b}+64 \cA T^{-ab}\Big(\cF^{-}_{cb}-i\cG^{-}_{cb}\Big) \cD_{d}\cH_{a}{}^{cd}\nn\\
	\cL_{9}&=&\cP^{2}\Big(\cF^{-ab}-i\cG^{-ab}\Big)^{2}+256T^{-ab}T^{-cd}C_{abef}C_{cd}{}^{ef}+512T^{-ab}T^{-cd}D{}C_{abcd}\nn\\
	&& +256\cP D^{2} +128T^{-ab}T^{-cd} \Big(V_{ab}\Big)^{i}{}_{j} \Big(V_{cd}\Big)^{j}{}_{i} +64\cP DT^{-ab}\Big(\cF^{-}_{ab}-i\cG^{-}_{ab}\Big) \nn\\
	&& +64\cP{}T^{-ab}\Big(\cF^{-}_{cb}-i\cG^{-}_{cb}\Big) \cD_{d}\cH_{a}{}^{cd}+64\cP{} T^{-ab}C_{abcd}\Big(\cF^{-cd}-i\cG^{-cd}\Big) \nn\\
	&& +32\cP {}\cD^{c}\cH_{abc}\cD_{d}\cH^{abd} -128T^{-ab}T^{-cd} \cD^{f}\cH_{acf} \cD^{g}\cH_{bdg}\nn \\
	&& -512T^{-ab}T^{-cd}C_{cdeb} \cD_{f}\cH_{a}{}^{ef}\nn\\
	\cL_{10} &=&- \frac{1}{2} \hat{\cC} \cdot T^+ + \Big( \cF^{-ab} -i \cG^{-ab} \Big) \hat{\cC}^-_{ab} - \frac{1}{2} B \Big( \cF^{-ab} -i \cG^{-ab} \Big)^2 +\frac{1}{2}\bar{B}\left(\cF\cdot\cF+\cG\cdot\cG\right)\nn\\
	\cL_{11} &=& \hat{\cC}^-_{ab} \hat{\cC}^{-ab} + \frac{1}{4} B^2 \Big( \cF^{-ab} -i \cG^{-ab} \Big)^2  - B\Big( \cF^{-ab} -i \cG^{-ab} \Big) \hat{\cC}^-_{ab}- \frac{1}{2} \cZ_{ij} \cZ^{ij}\nn \\
	\cL_{12}^{I} &=& 2 \hat{\cC}^{-ab} \hat{\cF}_{ab}^{-I} - X^I \Big( \cF^{-ab} -i \cG^{-ab} \Big) \hat{\cC}^-_{ab} + \frac{1}{2} X^I B \Big( \cF^{-ab} -i \cG^{-ab} \Big)^2  -\cZ_{ij}Y^{Iij} \nn \\
	&&   - B \Big( \cF^{-ab} -i \cG^{-ab} \Big) \hat{\cF}^{-I}_{ab} \nn\\
	\cL_{13} &=&   -2 A^\a Y_{ij}^\b Y^{ij} a_{\a \b} + B \cA \Big( \cF^{-ab} -i \cG^{-ab} \Big)^2 + 4 A^\a \hat{\cB}^{-\b}_{ab} \hat{\cC}^{-ab} a_{\a \b}  - 2 \cA \Big( \cF^{-ab} -i \cG^{-ab} \Big) \hat{\cC}^-_{ab} \nn \\
	&& - 2 BA^\a \Big( \cF^{-ab} -i \cG^{-ab} \Big) \hat{\cB}^{-\b}_{ab} a_{\a \b}\nn\\
	\cL_{14} &=&    -  32 T^{-ab} {C}^{cd}{}_{ab} \hat{\cC}^{-}_{cd}-32D\hat{\cC}^{-}_{ab}T^{-ab}-32T^{-ab}\hat{\cC}^{-}_{cb}\cD_{d}\cH_{a}{}^{cd} - 16  T^{-ab} \varepsilon^{ij} \Big(V_{ab}\Big)^k{}_i \cZ_{j k}\nn\\
	&&+32 B D \Big( \cF^{-ab} -i \cG^{-ab} \Big) T^-_{ab}  + \cP B  \Big( \cF^{-ab} -i \cG^{-ab} \Big) ^2 -2\cP\left(\cF^{-ab}-i\cG^{-ab}\right)\hat{\cC}^{-}_{ab}\nn\\
	&&  + 32 B \Big( \cF^{-cd} -i \cG^{-cd} \Big) T^{-ab} C_{abcd}+ 32 B \Big( \cF^{-}_{cb} -i \cG^{-}_{cb} \Big) T^{-ab} \cD_{d}\cH_{a}{}^{cd} 
	\end{eqnarray}
In arriving at the above action we have used the fact that once we impose the Poincar{\'e} gauge fixing condition, the dependent $U(1)$ R-symmetry gauge field (\ref{dependent}) is given in (\ref{U1}) and the associated field strength is given as:
\begin{align}\label{U1PGF}
R(A)_{ab}&=\frac{i}{4}\varepsilon_{abcd}\cD_{f}\cH^{cdf}
\end{align}
	\subsection{Final action}
	In this section, we will use the map between the Yang-Mills multiplet and the dilaton Weyl multiplet given in section-\ref{map} in the above action (\ref{fl}) and we will finally obtain the coupling of an arbitrary number of vector multiplets to $N=2$ Poincar{\'e} supergravity in the presence of arbitrary curvature squared terms. Its purely bosonic part is given as
	
	\begin{eqnarray}\label{FA}
	\cL&=&4\cD^a {G}_{I} \cD_a \bar{X}^I+\frac{1}{3}\varepsilon_{abcd}\cH^{bcd}G_{I}\overleftrightarrow{\cD^{a}}\bar{X}^{I} + 4 \cD^{a}({G}_{\cA} K^{ab}) \cD_a \bar{K}_{ab}  +\frac{1}{3}\varepsilon_{abcd}\cH^{bcd}(G_{\cA}K^{ef})\overleftrightarrow{\cD^{a}}K_{ef} \nn\\
	&&+ 32\cD_{a}(G_\cP T^{- ab}) \cD^c T^{+}_{cb}-8\cH^{bgh}(G_{\cP}T^{-}_{gh})\overleftrightarrow{\cD_{c}}T^{+c}{}_{b}  + 4\cD^a {G}_{B} \cD_a \bar{B}+\frac{1}{3}\varepsilon_{abcd}\cH^{bcd}G_{B}\overleftrightarrow{\cD}^{a}\bar{B}
\nn\\ 
&&+G\cL_{0}+G_{I}\cL_{1}^{I}+G_{\cA}\cL_{2}+G_{\cP}\cL_{3}+G_{\cA I}\cL_{4}^{I}+G_{\cP I}\cL_5^{I}+G_{IJ}\cL_6^{IJ}+G_{\cA\cA}\cL_{7}+G_{\cA\cP}\cL_{8}\nn\\
&& +G_{\cP\cP}\cL_{9}+ G_{B} \cL_{10} + G_{BB} \cL_{11} + G_{BI} \cL_{12}^{I} + G_{B \cA} \cL_{13}  +  G_{B \cP} \cL_{14} + \text{h.c.}
	\end{eqnarray}
	where,
	\begin{eqnarray}\label{FA1}
	\cL_{0}&=&\frac{1}{2}\Big(\cF^{-ab}-i\cG^{-ab}\Big)^2 \nn \\
	\cL_{1}^{I}&=&-\frac{1}{2}\hat{\cF}^{I}\cdot T^{+}+\frac{1}{2}\bar{X}^{I}\Big(\cF\cdot\cF+\cG\cdot\cG\Big)-\frac{1}{2}X^{I}\Big(\cF^{-ab}-i\cG^{-ab}\Big)^2+\Big(\cF^{-ab}-i\cG^{-ab}\Big)\hat{\cF}^{-I}_{ab}\nn\\
	\cL_{2}&=& \frac{1}{2} K^{ab} \bar{K}_{ab} \Big(\cF\cdot\cF+\cG\cdot\cG\Big) - 2 K^{ab} T^{+ cd} R_{cdab}(\o_-) -\frac{1}{4} K^2 \Big(\cF^{-ab}-i\cG^{-ab}\Big)^2  \nn\\
	&&  + 16 R^-_{abcd}(\o_-) R^{abcd}(\o_-)  + 8 \Big(V^{ab} \Big)^i{}_j \Big(V_{ab} \Big)^j{}_i  \nn \\
	\cL_{3}&=&-\frac{1}{2}\cP \Big(\cF^{-ab}-i\cG^{-ab}\Big)^2+128C_{abcd}C^{-abcd}+384D^{2}+64 \Big(V^{ab}\Big)^{i}{}_{j} \Big(V_{ab}\Big)^{-j}{}_{i}\nn\\
	&& +4\cH_{aef}\cH_{c}{}^{ef}T^{-ab}T^{+c}{}_{b} +16 R_{ab}T^{-ac}T^{+b}{}_{c} -32 \cD_{d}\cH^{abd} \cD^{e}\cH_{abe}
	\nn \\
	&&+16\varepsilon_{abcd} \cD_{f}\cH^{abf} \cD_{g}\cH^{cdg}
	\nn\\
	\cL_{4}^{I}&=& \frac{1}{2} K^2 X^{I}\Big(\cF^{-ab}-i\cG^{-ab}\Big)^2+ 8 K^{ab}\hat{\cF}^{-Icd}R^-_{cdab}(\o_-) - K^2\Big(\cF^{-ab}-i\cG^{-ab}\Big)\hat{\cF}^{-I}_{ab}    \nn\\
	&& -4 X^I K^{ab}\Big(\cF^{-cd}-i\cG^{-cd}\Big) R^-_{cdab}(\o_-) - 4 \varepsilon_{ik} K^{ab} \Big(V_{ab} \Big)^{k}{}_{j} Y^{I ij} 
	\nn \\
	\cL_{5}^{I}&=& \cP X^{I}\Big(\cF^{-ab}-i\cG^{-ab}\Big)^2-2 \cP \Big(\cF^{-ab}-i\cG^{-ab}\Big)\hat{\cF}^{-I} +32X^I DT^{-ab}\Big(\cF^{-}_{ab}-i\cG^{-}_{ab}\Big)\nn \\
	&& +32X^I T^{-ab}C_{abcd}\Big(\cF^{-cd}-i\cG^{-cd}\Big) -32T^{-ab}C_{abcd}\hat{\cF}^{-Icd}  -32DT^{-ab}\hat{\cF}_{ab}^{-I} \nn\\
	&&-16T^{-ab}\varepsilon^{ij}\Big(V_{ab}\Big)^{k}{}_{i}Y_{jk}^{I}-32T^{-ab}\hat{\cF}^{-I}_{cb} \cD_{d}\cH_{a}{}^{cd}+32 X^I T^{-ab}\Big(\cF^{-}_{cb}-i\cG^{-}_{cb}\Big)\cD_{d}\cH_{a}{}^{cd} 
	\nn \\
	\cL_{6}^{IJ}&=& \hat{\cF}^{-I}_{ab}\hat{\cF}^{-abJ}+\frac{1}{4}X^{I}X^{J}\Big(\cF^{-ab}-i\cG^{-ab}\Big)^{2}-\Big(\cF^{-ab}-i\cG^{-ab}\Big)X^{(I}\cF^{-J)}_{ab}-\frac{1}{2}Y^{I}_{ij}Y^{Jij} 
	\nn \\
	\cL_{7}&=& 16 K^{ab} K^{cd} R^-_{efcd} (\o_-) R^{-ef}{}_{ab} (\o_-) -4 K^2 K^{ab} \Big( \cF^{-cd} - i\cG^{-cd} \Big) R^-_{cdab}(\o_-) \nn\\
	&& + \frac{1}{4}\Big(K^2\Big)^2\Big(\cF^{-ab}-i\cG^{-ab}\Big)^{2} \nn\\ 
	\cL_{8}&=& K^2 \cP \Big(\cF^{-ab}-i\cG^{-ab}\Big)^{2} - 8 \cP K^{ab}\Big(\cF^{-cd}-i\cG^{-cd}\Big)R^- _{cdab} (\o_-)+ 32 K^2DT^{-ab}\Big(\cF^{-}_{ab}-i\cG^{-}_{ab}\Big) \nn \\
	&& -128 T^{-ab}K^{cd}C_{abef} R^{-ef}{}_{cd}(\o_-)   -128 DK^{ab}T^{-cd} R^-_{cdab}(\o_-) + 64K^{ab}T^{-cd} \big( V_{cd} \big)^j{}_i \big(V_{ab}\big)^i{}_j  \nn \\
	&&+32K^2T^{-ab}C_{abcd}\Big(\cF^{-cd}-i\cG^{-cd}\Big) -128T^{-ab}K^{ef} R^-_{cbef}(\o_-)\cD_{d}\cH_{a}{}^{cd}  \nn\\
	&&  +32 K^{2}T^{-ab}\Big(\cF^{-}_{cb}-i\cG^{-}_{cb}\Big)\cD_{d}\cH_{a}{}^{cd}
	 \nn \\
	\cL_{9}&=& \cP^{2}\Big(\cF^{-ab}-i\cG^{-ab}\Big)^{2}+256T^{-ab}T^{-cd}C_{abef}C_{cd}{}^{ef}+512T^{-ab}T^{-cd}DC_{abcd}\nn\\
	&&+256\cP D^{2} +128T^{-ab}T^{-cd} (V_{ab})^{i}{}_{j} (V_{cd})^{j}{}_{i} +64\cP DT^{-ab}\Big(\cF^{-}_{ab}-i\cG^{-}_{ab}\Big)
	\nn \\
	&& +64\cP T^{-ab}C_{abcd}\Big(\cF^{-cd}-i\cG^{-cd}\Big)  +64\cP T^{-ab}\Big(\cF^{-}_{cb}-i\cG^{-}_{cb}\Big)\cD_{d}\cH_{a}{}^{cd}\nn\\
	&&+32\cP \cD^{c}\cH_{abc}\cD_{d}\cH^{abd} -128T^{-ab}T^{-cd} \cD^{f}\cH_{acf}\cD^{g}\cH_{bdg}
	\nn\\	&& -512T^{-ab}T^{-cd}C_{cdeb}\cD_{f}\cH_{a}{}^{ef} 
	\nn\\
	\cL_{10} &=&- \frac{1}{2} \hat{\cC} \cdot T^+ + \Big( \cF^{-ab} -i \cG^{-ab} \Big) \hat{\cC}^-_{ab} - \frac{1}{2} B \Big( \cF^{-ab} -i \cG^{-ab} \Big)^2  +\frac{1}{2}\bar{B}\left(\cF\cdot\cF+\cG\cdot\cG\right)
	\nn\\
	\cL_{11} &=&  \hat{\cC}^-_{ab} \hat{\cC}^{-ab} + \frac{1}{4} B^2 \Big( \cF^{-ab} -i \cG^{-ab} \Big)^2  - B\Big( \cF^{-ab} -i \cG^{-ab} \Big) \hat{\cC}^-_{ab} - \frac{1}{2} \cZ_{ij} \cZ^{ij}
	\nn \\
	\cL_{12}^{I} &=&   2 \hat{\cC}^{-ab} \hat{\cF}_{ab}^{-I} - X^I \Big( \cF^{-ab} -i \cG^{-ab} \Big) \hat{\cC}^-_{ab} + \frac{1}{2} X^I B \Big( \cF^{-ab} -i \cG^{-ab} \Big)^2  -\cZ_{ij}Y^{Iij} \nn \\
	&&  - B \Big( \cF^{-ab} -i \cG^{-ab} \Big) \hat{\cF}^{-I}_{ab} 
	 \nn\\
	\cL_{13}&=&\frac{1}{2} K^2 B\Big(\cF^{-ab}-i\cG^{-ab}\Big)^2+ 8 K^{ab}\hat{\cC}^{-cd}R^-_{cdab}(\o_-)  - K^2\Big(\cF^{-ab}-i\cG^{-ab}\Big)\hat{\cC}^{-}_{ab} \nn\\
	&&  -4 B K^{ab}\Big(\cF^{-cd}-i\cG^{-cd}\Big) R^-_{cdab}(\o_-) - 4 \varepsilon_{ik} K^{ab} \Big(V_{ab} \Big)^{k}{}_{j} \cZ^{ij}
	 \nn \\
	\cL_{14} &=&  -  32 T^{-ab} {C}^{cd}{}_{ab} \hat{\cC}^{-}_{cd}-32D\hat{\cC}^{-}_{ab}T^{-ab}-32T^{-ab}\hat{\cC}^{-}_{cb}\cD_{d}\cH_{a}{}^{cd} - 16  T^{-ab} \varepsilon^{ij} \Big(V_{ab}\Big)^k{}_i \cZ_{j k} \nn\\
	&& +32 B D \Big( \cF^{-ab} -i \cG^{-ab} \Big) T^-_{ab}  + \cP B  \Big( \cF^{-ab} -i \cG^{-ab} \Big) ^2 -2\cP\left(\cF^{-ab}-i\cG^{-ab}\right)\hat{\cC}^{-}_{ab} \nn\\
	&& + 32 B \Big( \cF^{-cd} -i \cG^{-cd} \Big) T^{-ab} C_{abcd}+ 32 B \Big( \cF^{-}_{cb} -i \cG^{-}_{cb} \Big) T^{-ab} \cD_{d}\cH_{a}{}^{cd}
	\end{eqnarray}
	
	\noindent 
	While writing the above Lagrangian, we have introduced  the following notations for simplicity
	\begin{align}
	K^2 = K_{ab}K^{ab} \;\;\;,\;\;\;(K^2)^2 = (K_{ab}K^{ab})^2
	\end{align}
	Using 
	\begin{align}
	R_{abcd}(\o_-) = R_{abcd}(\o) - \cD_{[a}(\o) \cH_{b]cd} + \frac{1}{2} \cH_{[a|c}{}^e \cH_{|b] ed}
	\end{align}
	we can also extract the terms containing pure Riemann tensor from the above action. We can see from (\ref{FA},\ref{FA1}) that the terms containing Riemann tensor square, a combination of Weyl tensor square and Ricci scalar square and pure Ricci scalar square are contained in $\cL_{2}$, $\cL_{3}$ and $\cL_{11}$:
	\begin{align}
	\cL_{2}&=16 R^{-}_{abcd}R^{abcd}+\cdots\nn\\
	\cL_{3}&= 128 C_{abcd}C^{-abcd}+384 D^2+\cdots=128(C_{abcd}C^{-abcd}+\frac{R^2}{12})+\cdots\nn\\
	\cL_{11}&=-\frac{1}{2}\cZ_{ij}\cZ^{ij}+\cdots=-\frac{R^2}{8}+\cdots
	\end{align}
In the second and third line we have used the expression for the composite expressions for $D$ (\ref{T}) and $\cZ_{ij}$ (\ref{m}) upon Poincar{\'e} gauge fixing. On the other hand $\cL_{2}$, $\cL_{3}$ and $\cL_{11}$ come with $G_{\cA}$, $G_{\cP}$ and $G_{BB}$ respectively in the Lagrangian density. Hence by appropriately choosing the prepotential $G$ one can get supersymmetrization of arbitrary curvature square terms. For instance if one chooses the prepotential as shown below with the minimalistic set of terms required for getting curvature squared terms:
\begin{align}
G(X^{I},B,\cA,\cP)&=G^{(0)}(X^I)+\left(\alpha\cA+\beta\cP+\gamma B^2\right)G^{(1)}(X^{I})
\end{align}
then the curvature square term in the action would appear as:
\begin{align}
e^{-1}\cL&=16\left(\alpha R^{-}_{abcd}R^{abcd}+8\beta C^{abcd}C^{-}_{abcd} +\left(\frac{2}{3}\beta-\frac{1}{128}\gamma\right)R^2\right)G^{(1)}(X^{I})+\text{h.c}
\end{align}
and hence by appropriately tuning the parameters $\alpha$, $\beta$ and $\gamma$ one would get supersymmetrization of arbitrary curvature squared terms coupled to an arbitrary holomorphic function of the vector multiplet scalars.
	
	\section{Summary and Future Directions}\label{conc}
In this paper we have constructed the supersymmetrization of arbitrary curvature squared terms coupled to an arbitrary number of vector multiplets in $N=2$ Poincar{\'e} supergravity. We used the dilaton Weyl multiplet for our purpose. It is fairly well known from the work of Bergshoeff, de Roo and de Wit \cite{Bergshoeff:1980is} that if one starts with the chiral background constructed out of the standard Weyl multiplet via the identification of $T^{-}\cdot T^{-}$ with the lowest component of the chiral multiplet and uses the chiral density formula then one gets the supersymmetrization of pure Weyl tensor square term. One can couple arbitrary number of vector multiplets with this chiral background and upon using one of the vector multiplet as a compensator for Poincar{\'e} gauge fixing, one obtains the supersymmetrization of Weyl square term coupled to an arbitrary number of vector multiplets in $N=2$ Poincar{\'e} supergravity. Since the standard Weyl multiplet and the dilaton Weyl multiplet are related via the mapping (\ref{T}), one expects this construction to go through for the dilaton Weyl multiplet. The only differences being:
\begin{itemize}
\item
The dilaton Weyl multiplet is sufficient for partly gauge fixing the additional symmetries in the superconformal theory and one does not need any extra vector multiplet in this case unlike the standard Weyl multiplet.
\item
The field $D$ in the case of dilaton Weyl multiplet is a composite field which becomes proportional to the Ricci scalar upon Poincar{\'e} gauge fixing and because of the presence of the term $D^2$ in the action, one gets supersymmetrization of a combination of Weyl square and the Ricci scalar square term in the action
\end{itemize}
The dilaton Weyl multiplet has the advantage that one can construct supersymmetrization of Riemann square term in Poincar{\'e} supergravity. This is because, upon Poincar{\'e} gauge fixing, their exists a mapping between a Yang-Mills multiplet and the dilaton Weyl multiplet as discussed in section-\ref{map}. On the other hand one can construct an action for the Yang-Mills multiplet in four dimensions via embedding it in a chiral multiplet and using the chiral density formula. Further, the mapping between the Yang-Mills multiplet and the dilaton Weyl multiplet allows us to construct supersymmetrization of pure Riemann square term in $N=2$ Poincar{\'e} supergravity. At this point we would like to stress the importance of gauge fixing in our formalism. Unless we gauge fix, the mapping between the Yang-Mills multiplet and the dilaton Weyl multiplet does not exist. Hence in the superconformal formalism the action has pieces which are in terms of the fields belonging to the Yang-Mills multiplet. It is only when we gauge fix and apply the Yang-Mills map we get the action in terms of the supergravity multiplet, in particular the Riemann square term.

In order to construct supersymmetrization of Ricci scalar square term, one can use the tensor multiplet and its embedding into a reduced chiral multiplet (\ref{m}). This allowed us to construct supersymmetrization of arbitrary curvature squared terms coupled to an arbitrary number of vector multiplets. In order to do this we start in the superconformal theory where we couple the vector multiplets to a generic chiral background constructed out of the dilaton Weyl multiplet, a Yang-Mills multiplet and a tensor multiplet. All this is encoded in a single prepotential $F(X^I,X,B,\cA,\cP)$ which is a holomorphic as well as homogenous function of degree $2$. Upon Poincar{\'e} gauge fixing and using the mapping between the Yang-Mills multiplet and the dilaton Weyl multiplet we get the desired action in $N=2$ Poincar{\'e} supergravity which is the supersymmetrization of arbitrary curvature squared terms coupled to an arbitrary number of vector multiplets.

It is worth mentioning that one can also construct the supersymmetrization of arbitrary curvature squared terms using the standard Weyl multiplet. The supersymmetrization of Weyl square term is well known from the work of Bergshoeff, de Roo and de Wit almost 30 years back \cite{Bergshoeff:1980is}. The supersymmetrization of Ricci scalar square was obtained using the tensor multiplet in \cite{Kuzenko:2015jxa} and is analogous to the construction of Ricci scalar square invariant done in this paper. The third curvature squared invariant which is a particular combination of Ricci tensor square and Ricci scalar square was constructed in \cite{Butter:2013lta} based on the logarithm of a conformal primary chiral superfield. A particular combination of the Weyl scalar invariant and this invariant gives rise to the Gauss Bonnet invariant and is related via dimensional reduction to the five dimensional minimal supergravity \cite{Hanaki:2006pj} containing the Weyl square term and the gauge-gravitational Chern-Simons terms. This was crucial in solving a long standing puzzle regarding the matching of the microscopic entropy of a class of four dimensional extremal non BPS black holes with the supergravity calculation. The reason for the matching is because when these class of black holes (BPS as well as non BPS) are uplifted to five dimensions, they have an $AdS_3$ factor in their near horizon geometry and hence their entropy is completely determined by the gauge-gravitational Chern-Simons term and anything related to it by supersymmetry \cite{David:2007ak,Kraus:2005vz,Kraus:2005zm}.

In five dimensional minimal supergravity, apart from the invariant containing the Weyl square and the Chern-Simons terms, the Riemann square as well as the Ricci scalar square invariant was also constructed using the dilaton Weyl multiplet and the linear multiplet \cite{Ozkan:2013nwa}. This leads to the supersymmetrization of arbitrary curvature squared invariant in five dimensional minimal supergravity which we believe is related to our action via dimensional reduction for a specific choice of the prepotential. In order to establish this we first need to find the off-shell map between the dilaton Weyl multiplet in five dimensions and the dilaton Weyl multiplet and other matter multiplets in four dimensions via dimensional reduction similar to what was done for standard Weyl multiplets in five and four dimensions in \cite{Banerjee:2011ts}. We leave this for a future work.

It was argued by Kraus and Larsen in \cite{Kraus:2005vz,Kraus:2005zm} using anomaly inflow arguments and by Sahoo, Sen and David in \cite{David:2007ak} using AdS-CFT based arguments that if the black holes have an $AdS_3$ factor in their near horizon geometry then the only higher derivative terms that contribute towards the macroscopic entropy for these black holes are the Chern-Simons terms and anything else related to it by supersymmetry. Any other higher derivative corrections to the action which are supersymmetric by themselves should not affect the entropy of either the BPS or the non BPS black holes. It would be worthwhile to check this non-renormalization theorem directly using our action since our action is the most general higher derivative action in four dimensions and contains the reduction of a five dimensional Chern-Simons term as a sub-sector.

Although these additional invariants would not contribute to the entropy of the class of 4-dimensional black holes that has an $AdS_3$ factor in their near horizon geometry when uplifted to 5 dimensions, we expect that they would contribute to the entropy of a certain class of five dimensional black holes which has an $AdS_2$ factor in their near horizon geometry instead of $AdS_3$ studied in \cite{Cvitan:2007hu} and help resolve the puzzles concerning the entropy of such black holes as mentioned in \cite{Cvitan:2007hu}. We leave this for a future work.

In appendix-\ref{SU2breaking} we have reviewed how we can use the tensor multiplet to gauge fix the $SU(2)$-R symmetry. But we could only gauge fix it partially and there is still an unbroken $U(1)$ that is left. In order to fix the $U(1)$ we would need an $N=2$ superconformal multiplet which has a real scalar field in the fundamental of $SU(2)$ R-symmetry. Upon gauge fixing the $SU(2)$ to $U(1)$, the two components of this scalar field would combine into a complex scalar field transforming non trivially under the unbroken $U(1)$ which can be used to gauge fix the $U(1)$ and one would obtain a Poincar{\'e} supergravity where the R-symmetry is completely broken. Such a multiplet indeed exists which is the $N=2$ relaxed hypermultiplet \cite{Hegde:2020gpt}. This multiplet has two complex scalars transforming in the fundamental of the $SU(2)$ R-symmetry. The real part of one of the complex scalar can be used to gauge fix the unbroken $U(1)$. We will address this in a future work.

One would obtain a minimal coupling of the vector multiplets to $N=2$ supergravity when we truncate the prepotential $F(X,X^I)$ to depend only on X (the complex scalar field belonging to the dilaton Weyl multiplet) and $X^I$ (the complex scalar fields of the $N_V$ vector multiplets). In such a case one would expect that the $N_V$ physical scalar fields $Z^I$ would parametrize a special K\"{a}hler manifold. This means that the homogenous coordinates $X(Z)$ and $X^I(Z)$ together with the corresponding first derivatives of the prepotential $F_X(Z)$ and $F_I(Z)$ are expected to form a holomorphic section of $\mathcal{L}\otimes\mathcal{H}$, where $\mathcal{L}$ is a line bundle and $\mathcal{H}$ is a flat symplectic bundle and the K\"{a}hler form is given in terms of a symplectic invariant combination of the section. This is what happens for the case of minimal coupling of vector multiplet to $N=2$ supergravity arrived at from the superconformal formalism using the standard Weyl multiplet and one would expect the structure to be present in the case of the vector multiplet couplings obtained from the dilaton Weyl multiplet. However there are some noticeable differences because of which the features of ``special geometry'' is not immediately visible. We give our brief comments and leave working out the details for a future work. 

The $U(1)$ R symmetry gauge field $A_{\mu}$, the field $T_{ab}$ and $D$ appears as auxiliary fields in the case of the standard Weyl multiplet and their elimination by using the equations of motion played a crucial role in getting the special geometric features of the scalar manifold. However in the case of dilaton Weyl multiplet these fields are not fundamental and are rather expressed as composites in terms of the other fields of the multiplet as given in equations (\ref{dependent}) and (\ref{T}). In particular as shown in eq (\ref{dependent}), the $U(1)$- R symmetry gauge field $A_{\mu}$ depends on the 3-form field strength $\mathcal{H}_{\mu\nu\rho}$ which in turn depends on the bare gauge fields $W_{\mu}$ and $\tilde{W}_{\mu}$ as shown in (\ref{3formfieldstrength}). As a result of which the action is not completely in terms of the field strengths $F_{\mu\nu}$ and $G_{\mu\nu}$ corresponding to the gauge fields $W_{\mu}$ and $\tilde{W}_{\mu}$. There are Chern-Simons like terms in the action of the form $W\wedge F\wedge \mathcal{S}$ and $\tilde{W}\wedge G\wedge \mathcal{S}$, where $\mathcal{S}$ is some 1-form constructed out of the scalar fields and their derivatives. As a result of which one does not have duality transformations on the field strengths $F_{\mu\nu}$ and $G_{\mu\nu}$. The duality transformations will only act on the field strengths coming from the vector multiplet $F^{I}_{\mu\nu}$. If we just look at the sector of the Lagrangian which contains $F^{I}_{\mu\nu}F^{J}{}^{\mu\nu}$, we would notice that the duality transformations forms a symplectic group if we demand a dual Lagrangian and the scalar fields $X^I$ and the corresponding first derivative of the prepotential $F_{J}$ would transform as a symplectic vector. However the field strengths $F^{I}_{\mu\nu}$ also couple to the field strengths $F_{\mu\nu}$ and $G_{\mu\nu}$ and this sector of the Lagrangian along with the sector of the Lagrangian containing $F_{\mu\nu}F^{\mu\nu}$, $G_{\mu\nu}G^{\mu\nu}$ along with the Chern-Simons like terms for $W_{\mu}$ and $\tilde{W}_{\mu}$, makes the symplectic structure non trivial. Our preliminary investigations suggests that the scalar manifold admits a K\"{a}hler-Hodge structure and we are currently investigating the symplectic structure. Hopefully we will report on the work soon. 
\section*{Acknowledgements}
We thank Subramanya Hegde and Franz Ciceri for discussions and useful comments. This work is supported by SERB grant CRG/2018/002373, Government of India.
	
	\appendix 
	\section{Notations}\label{A1}
	
	\subsection{List of covariant derivatives used in this work}
	
	\begin{itemize}
		\item $\cD_\m$ :  denotes the fully supercovariant derivative with respect to all the bosonic gauge transformations except special conformal transformations in conformal supergravity. 
		\item  $\cD'_\m$ : fully supercovariant derivative with respect to all the bosonic gauge transformations except special conformal transformation and $U(1)_R$ in conformal supergravity.
		\item $\cD_\m$ : We will use the same notation $\cD_{\mu}$ in Poincar{\'e} supergravity to denote fully supercovariant derivative with respect to all bosonic gauge transformations in Poincar\' e supergravity theory with $SU(2)$ unbroken. Note that there is no dilatation or special conformal transformation or $U(1)$-R symmetry in the Poincar{\'e} supergravity. Hence by definition these are all excluded from $\cD_{\mu}$ when used in the context of Poincar{\'e} supergravity theory.
		\item  $\cD_\m(\o_{\pm})$ : denotes the covariant derivative same as $\cD_\m$ defined  with the standard spin connection replaced by the torsionful spin connection $\o_{\pm}$ as defined in (\ref{tsc}).
		\item $\cD_\m(\o, \o_{\pm})$ : This covariant derivative typically acts on a spinor that carries a Lorentz vector index as well. The covariant derivative rotates the Lorentz vector index via the spin connection $\o$, and the torsionful spin connection $\o_{\pm}$  defined in  (\ref{tsc}) rotates the Lorentz spinor index .
		
	\end{itemize}
	\section{{Identities for prepotential}} \label{A}
	The function F must
	be holomorphic and homogenous of second  degree with respect to its arguments $X^
	I$, $X$, $B$, $\cP$ and $\cA$:
	\begin{align} \label{prp}
	F(\alpha X^I, \a X, \a B, \a^2 \cA, \a^2 \cP) = \a^2 F(X^I, X, B, \cA, \cP)
	\end{align}
	By differentiating the defining relation (\ref{prp}) with respect to $\a$, and setting $\a = 1$, we get
	\begin{align}
	X^I F_I + X F_X +BF_B+ + 2 \cA F_\cA + 2 \cP F_\cP = 2 F( X^I, X,B, \cA, \cP)
	\end{align}
	
The notation for the derivatives and higher derivatives are defined as follows: 
		
	\begin{eqnarray}\label{prepotder}
	F_{I} &:=&\frac{\partial}{\partial X^{I}} F(X^I, X,B, \cA,\cP),  \quad
	F_{I J} :=\frac{\partial}{\partial X^{I}} \frac{\partial}{\partial X^{J}} F(X^I, X,B, \cA,\cP), \quad \nn\\
	F_{X} &:=&\frac{\partial}{\partial X} F(X^I, X,B, \cA,\cP),  \quad
	F_{XX} :=\frac{\partial}{\partial X} \frac{\partial}{\partial X} F(X^I, X,B, \cA,\cP), \quad \nn\\
	F_{B} &:=&\frac{\partial}{\partial X} F(X^I, X,B, \cA,\cP),  \quad
	F_{BB} :=\frac{\partial}{\partial B} \frac{\partial}{\partial B} F(X^I, X,B, \cA,\cP), \quad \nn\\
	F_{\cA} &:=&\frac{\partial}{\partial \cA} F(X^I, X,B, \cA, \cP),  \quad
	F_{\cA \cA} :=\frac{\partial}{\partial \cA} \frac{\partial}{\partial \cA} F(X^I, X, B,\cA,\cP)\quad \nn\\
	F_{\cP} &:=&\frac{\partial}{\partial \cA} F(X^I, X,B, \cA, \cP),  \quad
	F_{\cP \cP} :=\frac{\partial}{\partial \cP} \frac{\partial}{\partial \cP} F(X^I, X,B, \cA, \cP).
	\end{eqnarray}

\noindent	
	On Gauge fixing $X=\bar{X} = 1$ and defining,
	\begin{align}
          F(X^I, X, B, \cA, \cP)_{|X=1} &\equiv G (X^I , B, \cA, \cP)
	\end{align}
	we get the following relations:
	\begin{align}
	X^I G_I +B G_B + (F_X)_{|_{X =1}} + 2\cA G_\cA + 2 \cP G_\cP  = 2 G (X^I , B, \cA, \cP) 
	\end{align}
	Further identities are found by taking derivatives with respect to $X, \;X^I, \; \cA$ or $\cP$ as shown below:
	
	\begin{eqnarray}
	(F_{XI})_{|_{X=1}} &=&  {G}_I - X ^J G_{IJ}-BG_{BI} - 2 \cA G_{\cA I} - 2 \cP G_{\cP I}\nn\\
	(F_{X\cA})_{|_{X=1}} &=& -X^I G_{I \cA}-BG_{B\cA} - 2 \cA G_{\cA \cA}  - 2 \cP G_{\cP \cA} \nn\\
	(F_{X\cP})_{|_{X=1}} &=& -X^I G_{I \cP} (X^I , \cA, \cP ) - BG_{B\cP}(X^I , \cA, \cP )-2 \cP G_{\cP \cP}  (X^I , \cA, \cP) - 2 \cA G_{\cP \cA} (X^I , \cA, \cP) \nn\\
	(F_{XX})_{|_{X=1}} &=& 2 G - 2 X^I G_I -2BG_{B}- 2 \cA G_\cA  - 2 \cP G_\cP  
	\nn \\ 
	&+&  X^I X^J G_{IJ} +2X^{I}BG_{BI}+B^2G_{BB}+ 4 \cA X^I G_{I\cA} + 4 \cP X^I G_{I\cP} +4\cA BG_{B\cA}+ 4 \cP B G_{B\cP} 
	\nn \\
	&+& 8 \cA \cP G_{\cP \cA} +4 \cA \cA G_{\cA \cA} +  4 \cP \cP G_{\cP \cP}
	\end{eqnarray}
	
\section{Breaking of $SU(2)$-R symmetry}\label{SU2breaking}
In section-\ref{4}, we learned how to break dilatation, $U(1)$-R symmetry, special conformal transformation and S-supersymmetry by imposing certain gauge conditions on the dilaton Weyl multiplet. We also saw how the supersymmetry transformation rules gets modified so that the gauge conditions remain invariant. But we did not break $SU(2)$-R symmetry and continued working with Poincar{\'e} supergravity with $SU(2)$- R symmetry intact. In this section, we will learn how to break $SU(2)$-R symmetry and how it affects the supergravity multiplet and supersymmetry transformation laws. In order to break $SU(2)$-R symmetry, the dilaton Weyl multiplet would not be sufficient. We would need another compensating multiplet and we take that to be the tensor multiplet. We have seen earlier that we also needed the tensor multiplet for a different purpose. The dilaton Weyl multiplet alone was not sufficient for writing down an Einstein-Hilbert term in the action and we needed a tensor multiplet for that. In terms of the tensor multiplet fields we choose the $SU(2)$-gauge fixing condition as:
\begin{align}\label{C.1}
L_{ij}=L\delta_{ij}
\end{align}
However the above gauge conditions cannot break $SU(2)$ completely. There is a $U(1)$ subgroup of the $SU(2)$ that remains unbroken which is given by the following parametrization:
\begin{align}\label{C.2}
&\Lambda^{1}{}_{1}=\Lambda^{2}{}_{2}=0\nonumber \\
&\Lambda^{1}{}_{2}=-\Lambda^{2}{}_{1}=\lambda
\end{align}
where $\lambda$ is real. The $SU(2)$ that is broken is parametrized as:
\begin{align}\label{C.3}
&\Lambda^{1}{}_{1}=-\Lambda^{2}{}_{2}=\theta\nonumber \\
&\Lambda^{1}{}_{2}=\Lambda^{2}{}_{1}=i\sigma
\end{align}
where $\theta$ and $\sigma$ are real. Let us denote the Poincar{\'e} supersymmetry transformation with $SU(2)$ unbroken as $\delta^{P}$ which is related to the superconformal Q-supersymmetry transformation as given in (\ref{4.7}). We will further denote the Poincar{\'e} supersymmetry transformation with $SU(2)$ broken as $\hat{\delta}^{P}$. Since the gauge condition (\ref{C.1}) is not invariant under $\delta^{P}$, it suggests that $\hat{\delta}^{P}$ would be related to $\delta^{P}$ by the following field dependent (broken) $SU(2)$ transformation:
\begin{align}\label{C.4}
\hat{\delta}^{P}=\delta^{P}+\delta_{SU(2)}(\theta,\sigma)
\end{align}
where
\begin{align}\label{C.5}
\theta&=\frac{1}{2L}\left(\bar{\epsilon}_{1}\phi_1-\bar{\epsilon}_{2}\phi_2+\bar{\epsilon}^{2}\phi^2-\bar{\epsilon}^{1}\phi^1\right)\nonumber \\
i\sigma&=\frac{1}{2L}\left(\bar{\epsilon}_{1}\phi_2+\bar{\epsilon}_{2}\phi_1-\bar{\epsilon}^{2}\phi^1-\bar{\epsilon}^{1}\phi^2\right)
\end{align}
Given any fundamental of $SU(2)$ in the supergravity multiplet, for example $\phi^{i}$ or $\phi_{i}$, it is easy to see that they transform in the following way under the unbroken $U(1)$:
\begin{align}\label{C.6}
&\delta\phi^{1}=\lambda\phi^{2}\;,\;\; \delta\phi_{1}=\lambda\phi_{2}\nonumber \\
&\delta\phi^{2}=-\lambda\phi^1\;,\;\; \delta\phi_2=-\lambda\phi_1\nonumber \\
\end{align}
and hence we can define:
\begin{align}\label{C.7}
&\phi^{+}=\frac{1}{\sqrt{2}}\left(\phi^1-i\phi^2\right)\;,\;\; \phi^{-}=\frac{1}{\sqrt{2}}\left(\phi^1+i\phi^2\right)\nonumber \\
&\phi_{+}=\frac{1}{\sqrt{2}}\left(\phi_1-i\phi_2\right)\;,\;\; \phi_{-}=\frac{1}{\sqrt{2}}\left(\phi_1+i\phi_2\right)
\end{align}
which transforms under the unbroken $U(1)$ by the weights mentioned in the superscript or subscript i.e
\begin{align}\label{C.8}
\delta \phi^{\pm}=\pm i\lambda\phi^{\pm}\;,\;\; \delta \phi_{\pm}=\pm i\lambda\phi_{\pm}
\end{align}
we can define the Dirac conjugates in the same way and one can check that they are related to the Dirac conjugates of the spinors defined in (\ref{C.7}) as:
\begin{align}\label{C.9}
\bar{\phi}^{\pm}=\overline{\phi_{\mp}}\;,\;\; \bar{\phi}_{\pm}=\overline{\phi^{\mp}}
\end{align}
Similarly we can define the corresponding quantities $\psi_{\mu}^{\pm}$, $\psi_{\mu \pm}$, $\epsilon_{\pm}$ and $\epsilon^{\pm}$ for the gravitino and the supersymmetry transformation parameter\footnote{For fermions the placement of the $U(1)$ charges in the subscript or superscript will basically denote its chirality as $-1$ or $+1$ respectively. For bosons it does not matter where the charges are placed and by convention it will always be placed in the superscript.}. Among the three components of the $SU(2)$ gauge field $V_{\mu}{}^{i}{}_{j}$, one will remain as a gauge field of the unbroken $U(1)$ and the other two will transform under the unbroken $U(1)$ covariantly with some $U(1)$ charge. If we define:
\begin{align}\label{C.10}
&V_{\mu}{}^{1}{}_{1}=-V_{\mu}{}^{2}{}_{2}=\mathcal{V}_{\mu}\;, \nonumber \\
&\frac{1}{2}\left(V_{\mu}{}^{1}{}_{2}+V_{\mu}{}^{2}{}_{1}\right)=\mathcal{W}_{\mu}\;,\nonumber \\
&-\frac{1}{4}\left(V_{\mu}{}^{1}{}_{2}-V_{\mu}{}^{2}{}_{1}\right)=\mathcal{U}_{\mu}
\end{align}
One can check that under the unbroken $U(1)$, we get:
\begin{align}\label{C.11}
\delta \mathcal{U}_{\mu}&=\partial_{\mu}\lambda\nonumber \\
\delta \left(\mathcal{V}_{\mu}+i\mathcal{W}_{\mu}\right)&=-2i\lambda\left(\mathcal{V}_{\mu}+i\mathcal{W}_{\mu}\right)
\end{align}
Hence, the combination $\left(\mathcal{V}_{\mu}+i\mathcal{W}_{\mu}\right)$ transforms with a ($-2$) charge under the unbroken $U(1)$. But it does not transform covariantly under supersymmetry because of the compensating (broken) $SU(2)$ transformation (\ref{C.5}). Thus one has to make some field redefinitions to get a field which transforms covariantly under supersymmetry and one can easily check that the following combination would transform covariantly under supersymmetry:
\begin{align}\label{C.12}
\rho_{a}^{--}&\equiv \mathcal{V}_{a}+i\mathcal{W}_{a}+\frac{1}{2L}\left(\bar{\psi}_{a1}\phi_{1}-\bar{\psi}_{a2}\phi_{2}+\bar{\psi}_{a}^{2}\phi^{2}-\bar{\psi}_{a}^{1}\phi^{1}\right)+\frac{i}{2L}\left(\bar{\psi}_{a1}\phi_{2}+\bar{\psi}_{a2}\phi_{1}-\bar{\psi}_{a}^{1}\phi^{2}-\bar{\psi}_{a}^{2}\phi^{1}\right)\nonumber \\
&=\mathcal{V}_{a}+i\mathcal{W}_{a}+\frac{1}{L}\left(\bar{\psi}_{a-}\phi_{-}-\bar{\psi}_{a}^{-}\phi^{-}\right)
\end{align}
Thus the supergravity multiplet with the $SU(2)$ R-symmetry broken to a $U(1)$ is a 32+32 components multiplet as shown in Table-\ref{Table-supergravity}:
\begin{table}[H]
\caption{Poincar\'e supergravity multiplet.}\label{Table-supergravity}
\begin{center}
  \begin{tabular}{ | p{2cm}|p{8cm}|p{2cm}|}
\hline
    Field  & Description & Off-shell components\\ \hline
    $e_{\mu}{}^{a}$ & vielbein & 6 \\ \hline
    $\rho_{a}^{--}$ & Complex Lorentz vector. Comes from the original $SU(2)$ R-symmetry gauge field & 8 \\ \hline
    $\mathcal{U}_{\mu}$ & Real Gauge field of the unbroken $U(1)$ part of the $SU(2)$ R-symmetry & 3 \\ \hline
    $W_{\mu}$ & Real $U(1)$ Gauge field, Part of the original dilaton Weyl multiplet & 3 \\ \hline
    $\tilde{W}_{\mu}$& Real $U(1)$ Gauge field, Part of the original dilaton Weyl multiplet  & 3 \\ \hline
    $B_{\mu\nu}$ & Real tensor gauge field, Part of the original dilaton Weyl multiplet  & 3 \\ \hline
    $L$ &Real scalar field, Part of the original tensor multiplet  &1 \\ \hline
    $E_{\mu\nu}$ &Real tensor gauge field, Part of the original tensor multiplet & 3 \\ \hline
    $G$ & Complex scalar field, Part of the original tensor multiplet  & 2 \\ \hline
    $\psi_{\mu}^{\pm}$ & $\gamma^5=+1$, Gravitino & 24 \\ \hline
    $\phi^{\pm}$ & $\gamma^5=+1$, Part of the original tensor multiplet  & 8 \\ \hline
  \end{tabular}
\end{center}
\end{table}
The Poincar{\'e} supersymmetry transformations (with $SU(2)$ broken to $U(1)$) of the above mentioned fields can be calculated and are given as:
\begin{align}\label{C.13}
\hat{\delta}^{P} e_{\mu}^{a}&=\bar{\epsilon}^{-}\gamma^a\psi_{\mu +}+\bar{\epsilon}^{+}\gamma^a\psi_{\mu -}+\text{h.c}\nonumber \\
\hat{\delta}^{P}\psi_{\mu}^{+}&=2\hat{\mathcal{D}}_{\mu}(\omega_+)\epsilon^{+}-\rho_{\mu}^{++}\epsilon^{-}-\frac{i}{2}\gamma^{\rho}\left(\mathcal{F}_{\rho\mu}+i\mathcal{G}_{\rho\mu}\right)\epsilon_{+}-\frac{1}{2L}\left(\bar{\epsilon}_{+}\gamma^a\psi_{\mu}^{-}-\bar{\psi}_{\mu+}\gamma^a\epsilon^-\right)\gamma_{a}\phi_{+}\nonumber \\
&-\frac{1}{2L}\left(\bar{\psi}_{\mu}^{+}\epsilon^{-}-\bar{\epsilon}^{+}\psi_{\mu}^{-}\right)\phi^{+}-\frac{1}{8L}\left(\bar{\epsilon}^{+}\gamma_{ab}\psi_{\mu}^{-}-\bar{\psi}_{\mu}^{+}\gamma_{ab}\epsilon^{-}\right)\gamma^{ab}\phi^{+}\nonumber \\
\hat{\delta}^{P}\psi_{\mu}^{-}&=2\hat{\mathcal{D}}_{\mu}(\omega_+)\epsilon^{-}+\rho_{\mu}^{--}\epsilon^{+}+\frac{i}{2}\gamma^{\rho}\left(\mathcal{F}_{\rho\mu}+i\mathcal{G}_{\rho\mu}\right)\epsilon_{-}-\frac{1}{2L}\left(\bar{\epsilon}_{-}\gamma^a\psi_{\mu}^{+}-\bar{\psi}_{\mu-}\gamma^a\epsilon^+\right)\gamma_{a}\phi_{-}\nonumber \\
&-\frac{1}{2L}\left(\bar{\psi}_{\mu}^{-}\epsilon^{+}-\bar{\epsilon}^{-}\psi_{\mu}^{+}\right)\phi^{-}-\frac{1}{8L}\left(\bar{\epsilon}^{-}\gamma_{ab}\psi_{\mu}^{+}-\bar{\psi}_{\mu}^{-}\gamma_{ab}\epsilon^{+}\right)\gamma^{ab}\phi^{-}\nonumber \\
\hat{\delta}^{P} L&=\bar{\epsilon}_{+}\phi_{-}+\bar{\epsilon}_{-}\phi_{+}+\bar{\epsilon}^{+}\phi^{-}+\bar{\epsilon}^{-}\phi^{+}\nonumber \\
\hat{\delta}^{P}\phi^{-}&=\hat{\slashed{D}}L\epsilon_{-}+L\gamma^a\rho_{a}^{--}\epsilon_{+}-i\slashed{E}\epsilon_{-}-G\epsilon^{-}-\frac{1}{6}L\gamma^a\varepsilon_{abcd}\mathcal{H}^{bcd}\epsilon_{-}+\frac{i}{4}L\gamma\cdot\left(\mathcal{F}-i\mathcal{G}\right)\epsilon^{-}\nonumber \\
& +\frac{1}{L}\left(\bar{\epsilon}_{-}\phi_{-}-\bar{\epsilon}^{-}\phi^{-}\right)\phi^{+}\nonumber \\
\hat{\delta}^{P}\phi^{+}&=\hat{\slashed{D}}L\epsilon_{+}-L\gamma^a\rho_{a}^{++}\epsilon_{-}+i\slashed{E}\epsilon_{+}-G\epsilon^{+}-\frac{1}{6}L\gamma^a\varepsilon_{abcd}\mathcal{H}^{bcd}\epsilon_{+}-\frac{i}{4}L\gamma\cdot\left(\mathcal{F}-i\mathcal{G}\right)\epsilon^{+}\nonumber \\
& +\frac{1}{L}\left(\bar{\epsilon}_{+}\phi_{+}-\bar{\epsilon}^{+}\phi^{+}\right)\phi^{-}\nonumber \\
\hat{\delta}^{P}U_{\mu}&=-\frac{i}{2}\bar{\epsilon}_{+}\gamma^{\rho}R_{\rho\mu}^{-}+\frac{i}{2}\bar{\epsilon}_{-}\gamma^{\rho}R_{\rho\mu}^{+}+\frac{i}{2L}\bar{\epsilon}_{+}\phi_{+}\rho_{\mu}^{--}+\frac{i}{2L}\bar{\epsilon}_{-}\phi_{-}\rho_{\mu}^{++}\nonumber \\
& -\frac{i}{12}\bar{\epsilon}_{+}\gamma^a\psi_{\mu}^{-}\varepsilon_{abcd}\mathcal{H}^{bcd}+\frac{i}{12}\bar{\epsilon}_{-}\gamma^a\psi_{\mu}^{+}\varepsilon_{abcd}\mathcal{H}^{bcd}-\frac{i}{2L^2}\bar{\epsilon}_{+}\phi_{+}\left(\bar{\psi}_{\mu-}\phi_{-}-\bar{\psi}_{\mu}^{-}\phi^{-}\right)\nonumber \\
& +\frac{i}{2L^2}\bar{\epsilon}_{-}\phi_{-}\left(\bar{\psi}_{\mu+}\phi_{+}-\bar{\psi}_{\mu}^{+}\phi^{+}\right)+\text{h.c}\nonumber \\
\hat{\delta}^{P}\rho_{a}^{--}&=2\bar{\epsilon}_{-}\gamma^bR_{ba}^{-}-2\bar{\epsilon}^{-}\gamma^bR_{ba-}-\frac{2}{L}\left(\bar{\epsilon}_{-}D_{a}\phi_{-}-\bar{\epsilon}^{-}D_{a}\phi^{-}\right)-\frac{1}{L}\rho_{a}^{--}\left(\bar{\epsilon}_{+}\phi_{-}+\bar{\epsilon}^{+}\phi^{-}\right)\nonumber \\
&\frac{i}{2L}\left(\bar{\epsilon}_{-}\gamma^b\phi^{-}\kappa_{ba}-\bar{\epsilon}^{-}\gamma^{b}\phi_{-}\bar{\kappa}_{ba}\right)\nonumber \\
\hat{\delta}^{P}G&=-2\bar{\epsilon}_{+}\hat{\slashed{D}}(\omega_+)\phi^{-}-2\bar{\epsilon}_{-}\hat{\slashed{D}}(\omega_+)\phi^{+}-\bar{\epsilon}_{+}\gamma^a\phi^{+}\rho_{a}^{--}+\bar{\epsilon}_{-}\gamma^a\phi^{-}\rho_{a}^{++}\nonumber\\
&-\frac{4L}{3}\bar{\epsilon}_{+}\gamma^{ab}R_{ab-}-\frac{4L}{3}\bar{\epsilon}_{-}\gamma^{ab}R_{ab+}\nonumber \\
\hat{\delta}^{P}E_{\mu\nu}&=\bar{\epsilon}^{+}\gamma_{\mu\nu}\phi^{-}-\bar{\epsilon}^{-}\gamma_{\mu\nu}\phi^{+}+2L\bar{\epsilon}^{+}\gamma_{[\mu}\psi_{\nu]-}-2L\bar{\epsilon}^{-}\gamma_{[\mu}\psi_{\nu]+}+\text{h.c}\nonumber \\
\delta W_{\mu}&=-2i\bar{\epsilon}^{+}\psi_{\mu}^{-}+2i\bar{\epsilon}^{-}\psi_{\mu}^{+}+\text{h.c}\nonumber \\
\delta \tilde{W}_{\mu}&=2\bar{\epsilon}^{-}\psi_{\mu}^{+}-2\bar{\epsilon}^{+}\psi_{\mu}^{-}+\text{h.c}\nonumber \\
\delta B_{\mu\nu}&=\frac{1}{2}W_{[\mu}\hat{\delta}^{P}W_{\nu]}+\frac{1}{2}\tilde{W}_{[\mu}\hat{\delta}^{P}\tilde{W}_{\nu]}+2\bar{\epsilon}^{+}\gamma_{[\mu}\psi_{\nu]-}+2\bar{\epsilon}^{-}\gamma_{[\mu}\psi_{\nu]+}+2\bar{\epsilon}_{-}\gamma_{[\mu}\psi_{\nu]}^{+}+2\bar{\epsilon}_{+}\gamma_{[\mu}\psi_{\nu]}^{-}
\end{align}
where
\begin{align}\label{C.14}
\hat{\mathcal{D}}_{\mu}(\omega)\epsilon^{\pm}&=\partial_{\mu}\epsilon^{\pm}-\frac{1}{4}\gamma\cdot\omega_{\mu}\epsilon^{\pm}\mp iU_{\mu}\epsilon^{\pm}\;,\nonumber \\
\end{align}
and $\hat{D}$ is fully supercovariant w.r.t supersymmetry and the $U(1)$ gauge transformations. $R_{ab}^{\pm}$ are the fully supercovariant field strength for the gravitino defined as:
\begin{align}\label{C.15}
R_{ab}^{+}&=2\hat{\mathcal{D}}_{[a}(\omega,\omega^{+})\psi_{b]}^{+}+\rho_{[b}^{++}\psi_{a]}^{-}+\frac{i}{2}\gamma^{c}K_{c[b}\psi_{a]+}+\frac{1}{2L}\bar{\psi}_{[a+}\gamma^{c}\psi_{b]-}\gamma_{c}\phi_{+}+\frac{1}{2L}\bar{\psi}_{[b}^{+}\psi_{a]}^{-}\phi^{+}\nonumber \\
& +\frac{1}{8L}\bar{\psi}_{[a}^{+}\gamma^{cd}\psi_{b]}^{-}\gamma_{cd}\phi^{+}\nonumber \\
R_{ab}^{-}&=2\hat{\mathcal{D}}_{[a}(\omega,\omega^{+})\psi_{b]}^{-}-\rho_{[b}^{--}\psi_{a]}^{+}-\frac{i}{2}\gamma^{c}K_{c[b}\psi_{a]-}+\frac{1}{2L}\bar{\psi}_{[a-}\gamma^{c}\psi_{b]+}\gamma_{c}\phi_{-}+\frac{1}{2L}\bar{\psi}_{[b}^{-}\psi_{a]}^{+}\phi^{-}\nonumber \\
& +\frac{1}{8L}\bar{\psi}_{[a}^{-}\gamma^{cd}\psi_{b]}^{+}\gamma_{cd}\phi^{-}
\end{align}


\bibliography{references}
\bibliographystyle{jhep}

\end{document}